\documentclass[twocolumn]{aastex701}
\usepackage{natbib}
\usepackage{amsmath}

\bibpunct{(}{)}{;}{a}{}{,}

\begin{document}

\title{DAO: A New and Public Non-Relativistic Reflection Model}

\author[0009-0008-3299-9185]{Yimin Huang}
\affiliation{Center for Astronomy and Astrophysics, Center for Field Theory and Particle Physics and\\Department of Physics, Fudan University, Shanghai 200438, China}
\email{huangym23@m.fudan.edu.cn}  

\author[0000-0003-2845-1009]{Honghui Liu}
\altaffiliation{honghui.liu@uni-tuebingen.de}
\affiliation{Institut f\"ur Astronomie und Astrophysik, Eberhard-Karls Universit\"at T\"ubingen, D-72076 T\"ubingen, Germany}
\email{honghui.liu@uni-tuebingen.de}  

\author[0000-0002-3180-9502]{Cosimo Bambi}
\altaffiliation{bambi@fudan.edu.cn}
\affiliation{Center for Astronomy and Astrophysics, Center for Field Theory and Particle Physics and\\Department of Physics, Fudan University, Shanghai 200438, China}
\affiliation{School of Humanities and Natural Sciences, New Uzbekistan University, Tashkent 100001, Uzbekistan}
\email{bambi@fudan.edu.cn} 

\author[0000-0002-5311-9078]{Adam Ingram}
\affiliation{School of Mathematics, Statistics, and Physics, Newcastle University, Newcastle upon Tyne NE1 7RU, UK}
\email{Adam.Ingram@newcastle.ac.uk}

\author[0000-0002-9639-4352]{Jiachen Jiang}
\affiliation{Department of Physics, University of Warwick, Gibbet Hill Road, Coventry CV4 7AL, UK.}
\email{}

\author[0000-0003-3626-9151]{Andrew Young}
\affiliation{H.H. Wills Physics Laboratory, Tyndall Avenue, Bristol BS8 1TL, UK}
\email{Andy.Young@bristol.ac.uk}

\author[0000-0003-0847-1299]{Zuobin Zhang}
\affiliation{Astrophysics, Department of Physics, University of Oxford, Keble Road, Oxford OX1 3RH, UK}
\email{zuobin.zhang@physics.ox.ac.uk}

\begin{abstract}

We present a new non-relativistic reflection model, \texttt{DAO}, designed to calculate reflection spectra in the rest frame of accretion disks in X-ray binaries and active galactic nuclei. Employing the mathematical formalism of the \texttt{xillver} model, \texttt{DAO} couples the \texttt{XSTAR} code, which treats atomic processes, with the Feautrier method for solving the radiative transfer equation. A key feature of \texttt{DAO} is that it is open source and allows users to specify arbitrary illuminating spectra, enabling applications across diverse physical conditions. We incorporate a high-temperature corrected cross section and an exact redistribution function to accurately treat Compton scattering, and we use the most complete and up-to-date atomic database available. We investigate the spectral dependence on key physical parameters and benchmark the results against the publicly available table models for the widely used \texttt{reflionx} and \texttt{xillver} codes.
\end{abstract}

\keywords{\uat{High energy astrophysics}{739}; \uat{	
X-ray binaries}{1811}; \uat{AGN}{16};  \uat{Atomic physics}{2063}; \uat{Radiative transfer}{1335}}

\section{Introduction}

The X-ray spectra of X-ray binary systems (XRBs) and active galactic nuclei (AGNs) are essential for understanding the process of accretion onto black holes (BHs). These spectra typically consist of several components: (1) thermal emission from a geometrically thin and optically thick accretion disk \citep{Shakura1973, 1973blho.conf..343N}; (2) non-thermal emission from a hot corona ($\sim$100 keV), whose spectrum follows a power-law–like shape with a high-energy cutoff \citep{Thorne1975ApJ...195L.101T, Shapiro1976ApJ...204..187S, Zdziarski1995ApJ...438L..63Z}; and (3) reprocessed coronal emission from the accretion disk, commonly referred to as the reflection component \citep{1995Natur.375..659T, Risaliti2013Natur.494..449R, Bambi2021SSRv..217...65B}. The reflection spectrum encodes information about the disk properties (e.g., density, ionization state, and elemental abundances) as well as the spacetime geometry near the central black hole \citep{1989MNRAS.238..729F, George1991MNRAS.249..352G, 2001MNRAS.327...10B, 2010ApJ...718..695G, 2010MNRAS.409.1534D,2019ApJ...874..135T,2019ApJ...875...56T,2021ApJ...913...79T}. Modeling the reflection spectrum and fitting it to the data is particularly useful for extracting this information \citep{Bambi2021SSRv..217...65B, Reynolds2021ARA&A..59..117R, 2023ApJ...951..145L, Liu2023ApJ...950....5L, Draghis2024ApJ...969...40D}.

Calculating the disk reflection spectrum involves two steps: (1) computing the reflection spectrum in the local rest frame by solving the radiative transfer equation, taking into account processes such as photoionization and Compton scattering \citep{George1991MNRAS.249..352G, 2010ApJ...718..695G}; and (2) integrating the local reflection spectrum from the inner to the outer disk while including all relativistic effects, such as light bending, Doppler shifts, and gravitational redshift \citep{1989MNRAS.238..729F, 2010MNRAS.409.1534D, 2025MNRAS.536.2594L, 2025ApJ...989..168H}. Integration can be done based on a standard pre-tabulated transfer function \citep[see][for more details]{2024arXiv240812262B}.

For the rest-frame reflection spectrum, \texttt{reflionx}\footnote{\url{https://github.com/honghui-liu/reflionx_tables}} \citep{1978ApJ...219..292R, 1993MNRAS.261...74R, 2001MNRAS.327...10B, 2005MNRAS.358..211R, 2007MNRAS.381.1697R} and \texttt{xillver}\footnote{\url{https://www.sternwarte.uni-erlangen.de/~dauser/research/relxill/}} \citep{2010ApJ...718..695G, 2011ApJ...731..131G, 2013ApJ...768..146G} are the most widely used models in X-ray reflection data analysis. Another code, \texttt{TITAN} \citep{2003A&A...407...13D}, is also used in the literature.
The main difference between them lies in the treatment of atomic physics and radiative transfer. \texttt{xillver} adopts the Feautrier method for solving the radiative transfer equation, allowing the calculation of emergent spectra at different viewing angles.
It also includes fluorescent lines that are missing in \texttt{reflionx} and \texttt{TITAN}, due to the powerful atomic database of \texttt{XSTAR} \citep{2001ApJS..133..221K}. In addition, \texttt{TITAN} only includes Compton heating and cooling in the energy balance, without accounting for Comptonization in the redistribution of photon energy \citep{2000A&A...357..823D}.
The currently available \texttt{reflionx} and \texttt{xillver} table models employ distinct approaches to Compton scattering. 
In the \texttt{reflionx} model, \cite{1978ApJ...219..292R,1993MNRAS.261...74R,2005MNRAS.358..211R} utilize the Fokker-Planck equation \citep{1957JETP....4..730K} to address multiply scattered photons, while applying escape probabilities for unscattered and once-scattered line photons. 
For once-scattered line photons, \cite{1978ApJ...219..292R} approximate the energy redistribution using a Gaussian redistribution function.
In the \texttt{xillver} model, \citet{2010ApJ...718..695G,2013ApJ...768..146G,2014ApJ...782...76G} employ a Gaussian redistribution function to account for the energy shift in Compton scattering for all photons.
Although \citet{2000ApJ...537..833N} demonstrated the validity of treating all photons via the Gaussian approximation, this approximation breaks down at high photon energies ($E_{\rm ph} \gg m_e c^2$) or high electron temperatures ($kT_e \gg E_{\rm ph}$).

Energy redistribution in the Compton scattering process should be described by the full quantum-mechanical treatment, and the electrons in hot plasma should follow a relativistic Maxwellian distribution \citep{1993AstL...19..262N}. For scattering of low-energy photons by a low-temperature electron population, both the Gaussian and exact redistribution functions maintain nearly symmetric profiles. In other cases, the energy-exchange probabilities of photons differ markedly between the two formulations (see Appendix~\ref{apd:redistribution function}). \cite{2020ApJ...897...67G} investigated this effect on the reflection spectrum, and implemented it onto the \texttt{xillver} code, but table models including this new feature have not yet been publicly released and the \texttt{xillver} source code is not currently publicly available. They compared the spectrum calculated with the exact Compton scattering kernel with the publicly available \texttt{xillver} table and the Monte Carlo model \texttt{pexrav} \citep{1995MNRAS.273..837M}. The results demonstrate that the commonly used approximations for Compton scattering can lead to significant deviations in the high-energy band.
 
In addition, \cite{2018A&A...614A..79P} reported a noticeable discrepancy above 30~keV when comparing \texttt{xillverCp} with \texttt{pexrav}/\texttt{pexriv}. They attributed this difference to the fact that \texttt{xillverCp} does not account for the Klein–Nishina effect, which effectively absorbs part of the energy of the incident photons at $E \gtrsim 50$~keV \citep{2003MNRAS.342..355Z}.

In this paper, we present a new non-relativistic reflection code, \texttt{DAO}. We employ the \texttt{xillver} implementation: employing the photoionization code {\tt XSTAR} to compute atomic processes, thermal equilibrium, and ionization balance, and incorporating the Feautrier method to solve the radiative transfer. We include the exact redistribution function for Compton scattering provided by \cite{2020ApJ...897...67G}. We adopt a constant-density gas with a specified column density under a plane-parallel geometry and azimuthal symmetry. A key benefit of \texttt{DAO} is that it is open source and designed to be highly flexible, allowing users to adjust nearly all parameters. This will enable any user to, for example, create model grids for any arbitrary illuminating spectrum, or to implement more advanced model assumptions. A secondary benefit is the insight that can be gained from comparing the outputs of \texttt{DAO} and \texttt{xillver}. Since the mathematical formalism is identical, the two models should in principle agree exactly, although in practice we use different versions of ingredients such as {\tt XSTAR}. \texttt{DAO} is available on GitHub\footnote{\href{https://github.com/ABHModels/DAO.git}{https://github.com/ABHModels/DAO}} and Zenodo \citep{huang_2025_17839597}.

The manuscript is organized as follows. In Section~\ref{sec:METHOD}, we introduce the methodology of our model (radiative transfer, iteration process, etc.). The default configuration, as detailed in Section~\ref{sec:Basic setup}, is adopted throughout this work unless otherwise specified. In Section~\ref{sec:result}, we present a comparative analysis of our model against \texttt{reflionx} and \texttt{xillver}. We also investigate the impact of various parameters, such as the ionization parameter, incident spectrum, incidence angle, and hydrogen density, on the final reflection spectrum. Finally, Section~\ref{sec:discuess} summarizes our findings and outlines future research directions.

\section{METHOD}\label{sec:METHOD}
\begin{table*}
    \centering
    \begin{tabular}{c|l|c|c}
    \hline
    \hline
         Name & Description & Default & Unit \\
         \hline
         \multicolumn{4}{c}{\textbf{Grids and maximum iteration steps}}\\
         \hline
          N$_{\text{main}}$ & Maximum steps of main iteration loop & 15 & - \\
          N$_{\text{RTE}}$  & Maximum steps of radiative transfer iteration loop & 100 & - \\ 
          $N_\tau$ & Number of depth grid points & 200 & - \\
          $N_E$ & Number of energy grid points & 5000 & - \\
          $N_\mu$  & Number of angle grid points & 10 & - \\
          \hline
          \multicolumn{4}{c}{\textbf{Plasma}} \\
          \hline
          $\log\xi$ & Ionization parameter & 3.0 & erg cm s$^{-1}$ \\
          $n_h$ & Hydrogen density & $10^{15}$ & cm$^{-3}$ \\
          $A_{\text{Fe}}$ & Iron abundance & Solar & - \\
          $T_i$ & Initial gas temperature & $10^6$ & K \\
          $N_{iT}$ & Iterations for thermal equilibrium & 99 & - \\
          \hline
          \multicolumn{4}{c}{\textbf{Illumination}} \\
          \hline
          $I_{\text{type}}$ & Incident radiation field type & Power-law & - \\
          $\Gamma$/$kT_\mathrm{bb}^{\mathrm{top}}$ & Photon index / Blackbody temperature & 2.0 & - / eV \\
          $E_{\text{cut}}$/$kT_E$ & High energy cutoff / Electron temperature (nthcomp) & $3\times10^5$ & eV \\
          $kT_{\text{seed}}$ & Seed photon temperature (nthcomp) & 0.1$^*$ & keV \\
          $\mu_{\text{in}}$ & Incident angle ($\cos\theta$) & 0.7 & - \\ 
          $B_{\text{sw}}$ & Bottom illumination switch & Off & - \\
          \texttt{frac} & Ratio of top flux to total flux & 1 & - \\
          $kT_{\text{bb}}$ & Bottom blackbody temperature & 350 & eV \\
          \hline
          \multicolumn{4}{c}{\textbf{File paths}}\\
          \hline
          $P_\mathrm{sca}$ & Redistribution function & Eq.~\ref{eq:exact} & - \\
          $P_{\text{inci}}$ & User-defined incident spectrum (if I$_{\text{type}}$ = file) & - & - \\
          $P_{\mathrm{atom}}$ & Atomic database & data & - \\
          $P_\mathrm{out}$ & Output directory & - & - \\
    \hline
    \hline
    \end{tabular}
    \caption{Model parameters and their units. An asterisk (*) denotes a fixed parameter.}
    \label{model parameters}
\end{table*}
\subsection{Basic setup}\label{sec:Basic setup}
We consider a constant density slab illuminated from above and below. The slab represents the disk atmosphere, and the illumination from above represents emission from the corona irradiating the disk atmosphere, at an incidence angle $\theta_i$. \texttt{DAO} provides several standard forms for the incident spectrum, including a power law, a power law with a high-energy cutoff, the thermal Comptonisation model \texttt{nthComp} \citep{1996MNRAS.283..193Z}, and a blackbody. For the exponentially cut off power law, the specific intensity is given by\footnote{Note that our standard power law form does not include the 0.1 keV cut off that is adopted in \texttt{xillver}.}
\begin{equation}
I(E) = A E^{1-\Gamma} \exp( - E / E_{\rm cut} )
\end{equation}
and the \texttt{nthcomp} spectrum is parameterised by  $\Gamma$, the seed photon temperature $kT_{seed}$ (hardwired to $0.1$ keV for a multi-temeprature seed photon spectrum), and the electron temperature $kT_E$. These spectral forms have been widely used in the analysis of data from accreting black holes and neutron stars. Physical processes such as returning radiation and Comptonization can further modify the shape of the illumination \citep{1999MNRAS.309..561Z, 2021ApJ...910...49R, 2022MNRAS.514.3965D,2024ApJ...965...66M, 2024ApJ...976..229M}. To enable broader exploration of physical scenarios, \texttt{DAO} also allows user-defined incident spectra. The incident spectrum at the upper boundary is normalised by the ionization parameter \citep{1969ApJ...156..943T}
\begin{equation}\label{eq:ionization par}
\xi = \frac{ 4 \pi F_x }{ n_h },
\end{equation}
where $F_x$ is the illuminating X-ray flux\footnote{Our model calculates $F_x$ in the range of 0.1-1000 keV, such as {\tt xillver} \citep{2022ApJ...926...13G} but different from \texttt{reflionx}, which defines $F_x$ from 0.0013 to 1000 keV} and $n_h$ is the hydrogen number density.

\texttt{DAO} includes the option to incorporate a blackbody radiation source at the lower boundary, corresponding to the illumination of the disk atmosphere by thermal emission from the disk midplane. This feature is crucial for XRBs, as their accretion disks can be sufficiently hot to produce strong ionizing radiation \citep{Liu2023ApJ...950....5L}. A similar boundary condition is implemented in \texttt{reflionx} \citep{2007MNRAS.381.1697R}. At present, such bottom illumination is currently not included in the \texttt{xillver} table model. Therefore, to facilitate a direct comparison, we set the lower boundary condition to zero in this work.

Table~\ref{model parameters} lists the input parameters of the \texttt{DAO} model. These parameters are grouped into four categories: (1) model grids and maximum iteration steps, (2) initial plasma conditions, (3) boundary illumination, and (4) required file paths.

The grid-related parameters define the discretization of energy, angle, and optical depth, as well as the maximum number of iteration steps. In this work, we adopt 10 linearly spaced cosine-angle bins ranging from $0.05$ to $0.95$ for the angular quadrature. For the remaining dimensions, a Thomson optical-depth grid consisting of 200 points logarithmically spaced from $10^{-4}$ to $10$ is used, together with a default energy grid of 5000 points covering the range from $0.1\,\mathrm{eV}$ to $1000\,\mathrm{keV}$. The maximum number of main iteration steps, $N_\mathrm{main}$, and radiative-transfer iteration steps are set to 15 and 200, respectively. These values are sufficient to ensure convergence of the model, as discussed in detail in Section~\ref{sec:equ}.

Regarding the plasma parameters, \texttt{DAO} assigns strictly non-zero default abundances to H, He, C, O, Ne, Mg, Si, S, Ar, Ca, Fe, and Ni. Nevertheless, the abundances of all elements from H to Zn can be freely modified by the user. In practice, we vary only the iron abundance, while keeping all other elemental abundances fixed at solar abundance. Consequently, only the iron abundance, $A_\mathrm{Fe}$, is listed in Table~\ref{model parameters}.

The file-path parameters specify the locations of the pre-calculated Compton scattering redistribution functions, user-defined incident spectra, the atomic database, and the output files. The atomic database employed in \texttt{DAO} is the same as that used by \texttt{XSTAR}\footnote{We use ATDB 2024 in this paper}. The output paths correspond to four files that store, respectively, the final temperature profile; ion populations and heating–cooling rates at each depth; the emergent intensity from $\mu_1$ to $\mu_\mathrm{N_\mu}$; and the mean intensity and illumination.

\subsection{Radiative transfer equation}

This section outlines the methodology employed in our reflection model \texttt{DAO}, which follows the mathematical formalism employed for the \texttt{xillver} models \citep{2010ApJ...718..695G,2013ApJ...768..146G}. We employ {\tt XSTAR}~v2.59 \citep{2021ApJ...908...94K} to treat the relevant atomic processes, and adopt the Feautrier method (hereafter FTM) \citep{1978stat.book.....M,2015tsaa.book.....H} to iteratively solve the radiative transfer equation.

We consider a plane-parallel atmosphere. Under this assumption, the radiative transfer equation can be written as:
\begin{equation}\label{eq:rte}
    \mu\frac{\partial I(z,\mu,E)}{\partial \tau(E)} = S(z,\mu,E)-I(z,\mu,E)
\end{equation}
where the cosine of the angle between the photon propagation direction and the disk normal, $\mu$, and the photon energy, $E$, define the grids that characterize the radiation field quantities: the specific intensity, $I$, and the source function, $S$. The monochromatic optical depth at energy $E$ is
\begin{equation}
    d\tau(E)=-\chi(E) dz
\end{equation}
where $z\,[\mathrm{cm}]$ denotes the height of each layer, and $\chi(E)\,[\mathrm{cm^{-1}}] = \alpha(E) + \alpha^c(E)$, with $\alpha^c$ and $\alpha$ being the scattering and absorption coefficients, respectively.

FTM is based on the solution of the second-order radiative transfer equation \citep{1905ApJ....21....1S}, which is derived by considering two rays propagating in opposite directions, denoted by $\pm\mu$. From these, we define the mean-like intensity at depth $z$ and in the range $0\le\mu\le 1$ as
\begin{equation}\label{eq:mean-like}
    u(z,\mu,E)= \frac{1}{2}\left[I(z,+\mu,E)+I(z,-\mu,E)\right],
\end{equation}
and the flux-like intensity as
\begin{equation}
    h(z,\mu,E) = \frac{1}{2}\left[I(z,+\mu,E)-I(z,-\mu,E)\right].
\end{equation}
By adding and subtracting Eq.~\ref{eq:rte} for $+\mu$ and $-\mu$, we obtain the transfer equations for $h$ and $u$. Substituting the equation for $u$ into that for $h$, using the Thomson optical depth ($\tau_T = \int n_e \sigma_T dz$, where $\sigma_T = 6.652\times10^{-25}\,\mathrm{cm}^2$ is the Thomson scattering cross section) to define the depth grid, and abbreviating the notation, we obtain the second-order radiative transfer equation:
\begin{equation}\label{eq:second-order_rte}
    \mu^{2} \frac{\partial^2 u(\mu,E)}{\partial\tau^{2}(E)} = u(\mu,E) - S(E)
\end{equation}

A key assumption in this equation is the forward--backward symmetry of the source function\footnote{In a real disk irradiated by a compact corona, the symmetry between forward and backward directions is broken because the radiation is strongly anisotropic.}, i.e., $S(+\mu) = S(-\mu)$. This symmetry is essential for solving the second-order radiative transfer equation (Eq.~\ref{eq:second-order_rte}) using the FTM. When this approximation is not valid, alternative methods must be employed \citep{1975ApJ...202..250M}. The source function can then be expressed as
\begin{equation}\label{eq:source_function}
    S(E) = \frac{j(E)}{\alpha^c(E)+\alpha(E)} + \frac{\alpha^c(E) J^c(E)}{\alpha^c(E)+\alpha(E)}
\end{equation}
where $j(E)$ represents the emissivity, incorporating contributions from thermal, line, and radiative recombination, $\alpha(E)$ includes free-free, bound-free, and radiative recombination opacity. \texttt{XSTAR} can directly compute both $j(E)$ and $\alpha(E)$ for a given ionizing flux, column density, and gas temperature. The scattering coefficient is given by $\alpha^c(E)= n_e\sigma_{cs}(E)$, where $\sigma_{cs}(E)$ is the energy-dependent Compton scattering cross section averaged over a relativistic Maxwellian electron distribution, and $n_e=1.2n_\mathrm{H}$ is the electron density.

The second term on the right-hand side in Eq.~\ref{eq:source_function} represents the Compton scattering of photons by electrons in the gas. Unlike the Klein--Nishina scattering cross section used in previous reflection models \citep{2005MNRAS.358..211R}, which is only valid for low-temperature electrons \citep[see Figure 1 in][]{2020ApJ...897...67G}, our approach accounts for high-temperature gas by employing the Compton scattering cross section averaged over a relativistic Maxwellian electron distribution \citep{1996ApJ...470..249P}. The mathematical form of $\sigma_{cs}$ can be found in Eq.~\ref{eq:rel_cs} in Appendix~\ref{apd:redistribution function}. $J^c$ is the mean intensity after scattering:
\begin{equation}\label{eq:scattered_mean_intensity}
    J^c (E_i) = \frac{1}{\sigma_{cs}(E_i)}\int_{E_{\rm min}}^{E_{\rm max}} R(E_f,E_i)J(E_f)dE_f
\end{equation}
where $E_i$ and $E_f$ are the photon energies before and after scattering, respectively, and $J$ is the mean intensity before scattering. The Compton scattering redistribution function, $R$, satisfies the detailed balance condition relating the initial and final photon energies, $E_i$ and $E_f$, respectively. This condition holds for a Maxwellian electron distribution and is consistent with the photons following a Wien distribution in thermal equilibrium \citep[see][Chapter VIII]{1973erh..book.....P}:
\begin{equation}
R(E_f, E_i) = R(E_i, E_f) \exp{\left(-\frac{E_f - E_i}{k_{\rm B} T_e}\right)}
\end{equation}
Note that the photon phase-space factors are implicitly absorbed into the definition of $R$. We adopt the normalization condition given by \citep{1996ApJ...470..249P,2017MNRAS.469.2032M}:
\begin{equation}
\frac{\sigma_{\rm cs}(E_i)}{\sigma_{\rm T}} = \int_0^\infty R(E_i, E_f) , dE_f
\end{equation}

The redistribution function $R$ governs the change in photon energy during the scattering process (see Appendix \ref{apd:redistribution function}). In addition to altering photon energies, Compton scattering also modifies photon propagation directions. At the present stage, we employ the angle-averaged redistribution function, consistent with the approaches used in \texttt{xillver} and \texttt{reflionx}. \cite{2012A&A...545A.120S} showed that there exists a $\sim 2\%$ difference between the angle-dependent and angle-averaged redistribution functions in a pure hydrogen atmosphere (see their Figure~2). In future work, we plan to incorporate the full angular dependence of directional changes.

Previous public reflection models have employed a Gaussian-approximated redistribution function. In this work, we compare \texttt{DAO} with other models using both the Gaussian approximation and a more rigorous second-order exact redistribution function previously explored by \cite{2020ApJ...897...67G} (see Section~\ref{Comparison with other model}). We further quantify the deviations introduced by the Gaussian approximation in Section~\ref{dif kernel}. The mathematical forms of both the second-order exact and Gaussian-approximated redistribution functions are provided in Appendix~\ref{apd:redistribution function}. To facilitate flexibility, \texttt{DAO} includes a built-in selector that allows users to switch between the two redistribution schemes.

\subsection{Boundary condition}

To solve the second-order radiative transfer equation (Eq.~\ref{eq:second-order_rte}), we again use the same mathematical formalism used by \texttt{xillver}, namely FTM. Specifically, we reformulate the radiative transfer equation as a block tridiagonal system of the form $T \cdot j = R$ (see Appendix~\ref{apd:algorithm} for a detailed derivation) and then solve with the Thomas algorithm, which performs recursive forward elimination followed by back substitution.

The disk surface is irradiated by hard X-ray photons from the corona, while the disk itself emits thermal radiation. This configuration can be formulated as a two-boundary problem, making it suitable for solving the radiative transfer equation (Eq.~\ref{eq:second-order_rte}). The upper boundary condition is determined by the incident radiation from an external source such as the corona. Following \citet{1967ApJ...150L..53A}, we adopt the second-order exact boundary condition derived from a Taylor series (see Appendix~\ref{apd:algorithm}):
\begin{equation}\label{eq:boundary}
    \mu \frac{u_2 - u_1}{\Delta\tau_{3/2}} = u_1 + \Delta\tau_{3/2} \frac{u_1 - S_1}{2\mu} + I_{\mathrm{inc}}(\mu_\mathrm{inc})
\end{equation}
where $u_1$ and $u_2$ are the mean-like intensities (Eq.~\ref{eq:mean-like}) at the first and second layers, respectively; $I_{\mathrm{inc}}$ represents the incident intensity at incidence angle $\mu_\mathrm{inc}$, originating from the X-ray source near the disk. For the lower boundary, when $B_\mathrm{sw}=1$, a blackbody is assumed as the disk radiation to illuminate the bottom surface. The second-order exact boundary condition at the bottom is given by
\begin{equation}\label{eq:lower_boundary}
    \mu \frac{u_{D}-u_{D-1}}{\Delta\tau_{D-1/2}} = B_D^+ - u_D - \Delta\tau_{D-1/2}\frac{\left(u_D-S_D\right)}{2\mu}
\end{equation}
where $D=ND$. When $B_\mathrm{sw}=0$, we have $B_D^+ = 0$. Equation~\ref{eq:dif from rte}, together with Eq.~\ref{eq:boundary} and Eq.~\ref{eq:lower_boundary}, forms a block tridiagonal system
\begin{equation}
    T \cdot j = R
\end{equation}
which is straightforward to solve by recursive forward elimination and back substitution.


\subsection{Equilibrium and iteration}\label{sec:equ}

Convergence must be achieved through an iterative procedure. We begin by assuming that the initial radiation field at each depth is equal to the coronal illumination. Then, we use \texttt{XSTAR} to calculate the emissivity and opacity at each depth (excluding scattering in this step). The \texttt{XSTAR} calculations enforce thermal equilibrium, satisfying
\begin{equation}\label{eq:H=c}
H = C ,
\end{equation}
where $H$ is the heating rate and $C$ is the cooling rate. Once the plasma reaches thermal equilibrium under the current radiation field, we assume that it remains in this equilibrium state during the radiative-transfer calculation. In other words, the plasma temperature, emissivity, and absorption coefficients are held fixed in this stage.

The solution to the radiative transfer equation is obtained through a standard $\Lambda$-iteration scheme as follows:
\begin{enumerate}
\item Using the current temperature, emissivity, and absorption, we compute the source function (Eq.~\ref{eq:source_function}) at each Thomson optical depth $\tau_T$.
\item We solve the second-order radiative transfer equation (Eq.~\ref{eq:second-order_rte}) using the Feautrier method (see Appendix \ref{apd:algorithm}).
\item We update the source function based on the new radiation field obtained at step~2 and use this new source function to update the radiation field.
\end{enumerate}
This iteration loop stops when the source function converges:
\begin{equation}\label{eq:conver}
    \int\frac{S^{n}(E)-S^{n-1}(E)}{S^{n-1}(E)}dE <e_n,\quad e_n = 10^{-10}
\end{equation}

After the radiation field converges in the radiative-transfer step, we again use this updated radiation field as the input to \texttt{XSTAR} to obtain a new plasma profile. We then solve the radiative transfer equation again. The iterative procedure continues until both the temperature and ionization profiles at each layer no longer change. When \texttt{DAO} calculates low-ionization models, photoionization is the dominant process, with $\alpha \gg \alpha^c$. In this regime, Eq.~\ref{eq:conver} converges more rapidly than in a scattering-dominated gas. In contrast, for high-ionization models, the gas is scattering-dominated, with $\alpha^c \gg \alpha$, leading to slower convergence. For example, Model~1 (Table~\ref{tab:computation time}) requires fewer than 50 steps per radiative transfer calculation to satisfy Eq.~\ref{eq:conver}, while Model~3 (Table~\ref{tab:computation time}) requires nearly 100 steps.

The total runtime of \texttt{DAO} strongly depends on the physical conditions and model settings. In cases with low ionization parameters and soft illumination, the gas reaches cooler temperatures, which requires {\tt XSTAR} to carry out more computationally intensive evaluations of radiative quantities at each optical depth. Consequently, the time per iteration of the main loop increases substantially. Our computations were performed on a server equipped with four sockets, each hosting an Intel(R) Xeon(R) CPU E7-8860 v3 processor. A typical run on a single core requires approximately 97 hours for the full set of iterations in Model 1 (Table~\ref{tab:computation time}), with each individual $\Lambda$-iteration taking about 30 minutes. In contrast, when the disk is illuminated by harder photons and set to a higher ionization parameter, as in Model 2 (Table~\ref{tab:computation time}), the total runtime decreases to 36 hours, while each $\Lambda$-iteration still requires roughly 30 minutes.

Considering that \texttt{XSTAR} is the most time-consuming component of the calculation and that we employ a high energy resolution, we do not use the Accelerated Lambda Iteration (ALI) method in this work. In our case, the accelerated $\Lambda^{*}$-operator becomes a full matrix of dimension $N_{\tau_T}\times N_E \times N_E$ because the partial redistribution function (specifically, the scattering redistribution function in our context) is considered \citep{1995A&A...297..771P}, making the corresponding linear system prohibitively expensive to solve. Moreover, \cite{2012A&A...545A.120S} demonstrated that ALI still requires a large number of iterations when the exact Compton scattering redistribution function is used

\begin{table} 
    \centering
    \begin{tabular}{c|c|c|c|c|c|c|c}
    \hline
         Model& N$_E$&N$_\tau$&N$_\mu$ &$\log\xi$ &Incident& $\Gamma$ & kT$_E$\\
    \hline
    \hline
    1& 5000 & 200 & 10 & 1.0 & \texttt{nthcomp} & 2.0 & 60 kev \\

    2& 5000 & 200 & 10 & 3.0 & \texttt{nthcomp} & 1.4 & 60 kev \\

    3 & 5000 & 200 & 10 & 3.0 & \texttt{nthcomp} &2.0 & 60 keV \\
    \hline
    \hline
    \end{tabular}
    \caption{Three different model sets for moderate nthcomp incident with low ionization degree gas (Model 1), hard nthcomp incident with high ionization degree gas (Model 2) and moderate nthcomp incident with high ionization degree gas (Model 3)}
    \label{tab:computation time}
\end{table}

\subsection{Implementation}
In summary, the \texttt{DAO} code utilizes the public \texttt{XSTAR} v2.59 \citep{2001ApJS..133..221K,2021ApJ...908...94K} to treat atomic physics. The differences in the treatment of atomic physics among \texttt{DAO}, \texttt{xillver}, and \texttt{reflionx} are summarized in Table \ref{Atom table version}. For coronal illumination, the cut-off power law, power law, blackbody, and custom-defined spectra are implemented. In addition, we incorporate public code from XSPEC \citep{1996ASPC...99..117G} \texttt{nthcomp} \citep{1996MNRAS.283..193Z,1999MNRAS.309..561Z} and \texttt{comptt} \citep{1994ApJ...434..570T} as illumination spectra. For Compton scattering, we use either a pre-calculated exact or Gaussian-approximated redistribution function. The exact redistribution function is calculated using the GitHub repository provided by \cite{2020ApJ...897...67G}. For radiative transfer, we employ the FTM, which has been widely applied in the scenario of X-ray reflection \citep{2000ApJ...537..833N,2010ApJ...718..695G,2013ApJ...768..146G}. Due to the proprietary nature of the \texttt{xillver} code, we can only compare our code with its public table models, which will be discussed in the next section.

\begin{figure*}
    \centering
    \includegraphics[width=\linewidth]{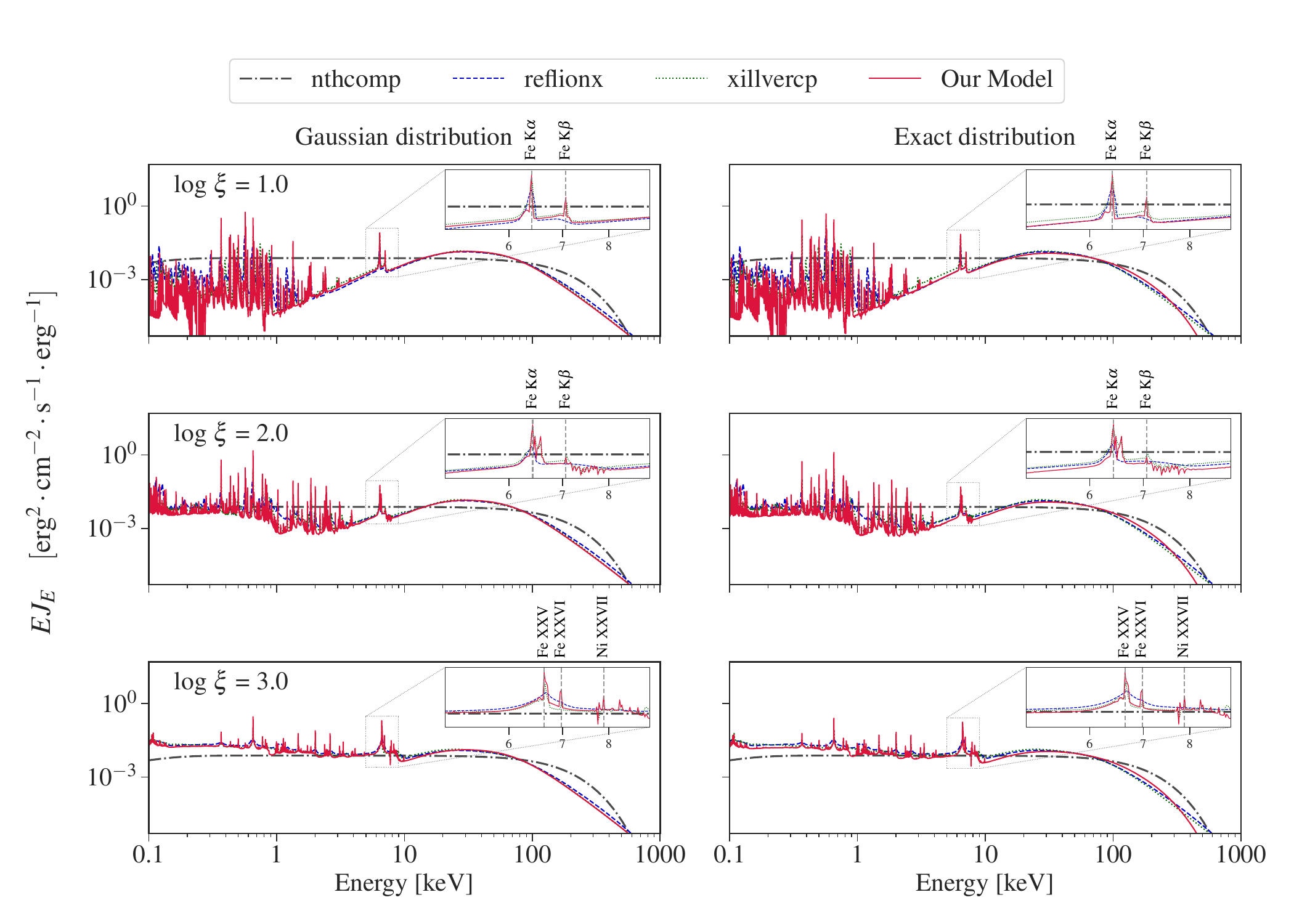}
    \caption{Reflection spectra generated by our model (\texttt{DAO}, red solid line), compared with those from \texttt{reflionx} (blue dashed line) and \texttt{xillvercp} (green dotted line). The incident \texttt{nthcomp} spectrum from \texttt{xillvercp} (\texttt{refl\_frac}=0, $\Gamma$ = 2, $kT_E$ = 60 keV) is depicted by the grey dash-dotted line, iron abundance is set at 1.32 for \texttt{DAO} and \texttt{xillver} to account for different abundance tables. Other parameters are set at default their value as shown in Table \ref{model parameters}. To compare fairly with \texttt{reflionx} and \texttt{xillver}, we also run \texttt{DAO} using the Gaussian approximation. Left column: Gaussian-approximated redistribution function. Right column: exact redistribution function. All spectra are normalized by their integral of energy, $EJ_\mathrm{norm}(E) = EJ(E)/\int EJ(E)dE$.}
    \label{Comparsion different model with nthcomp incident}
\end{figure*}
\begin{figure*}
    \centering
    \includegraphics[width=\linewidth]{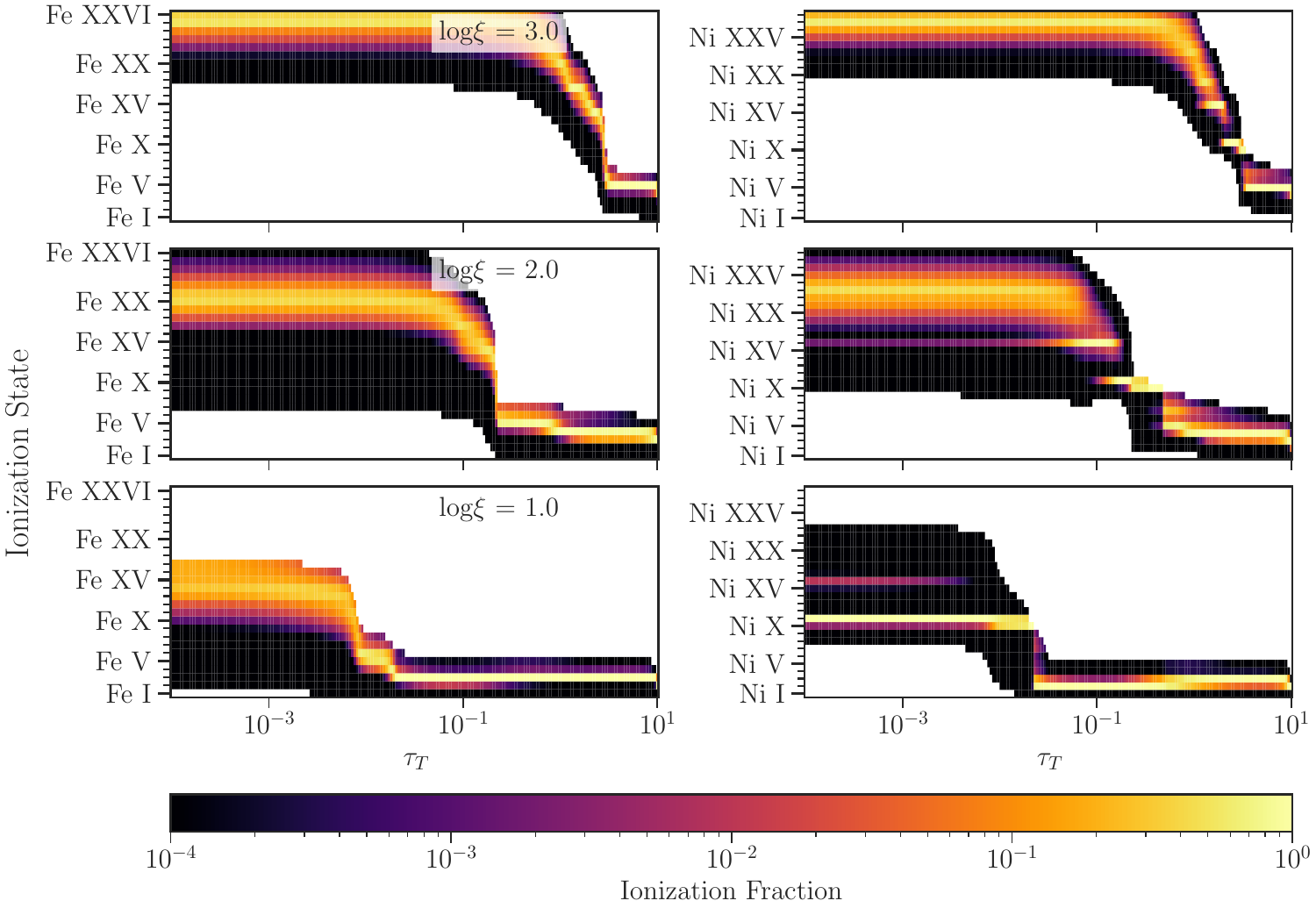}
    \caption{Ions fraction of Fe and Ni for reflection shown in Figure \ref{Comparsion different model with nthcomp incident}}
    \label{fig:ni}
\end{figure*}
\begin{figure*}
    \centering
    \includegraphics[width=\linewidth]{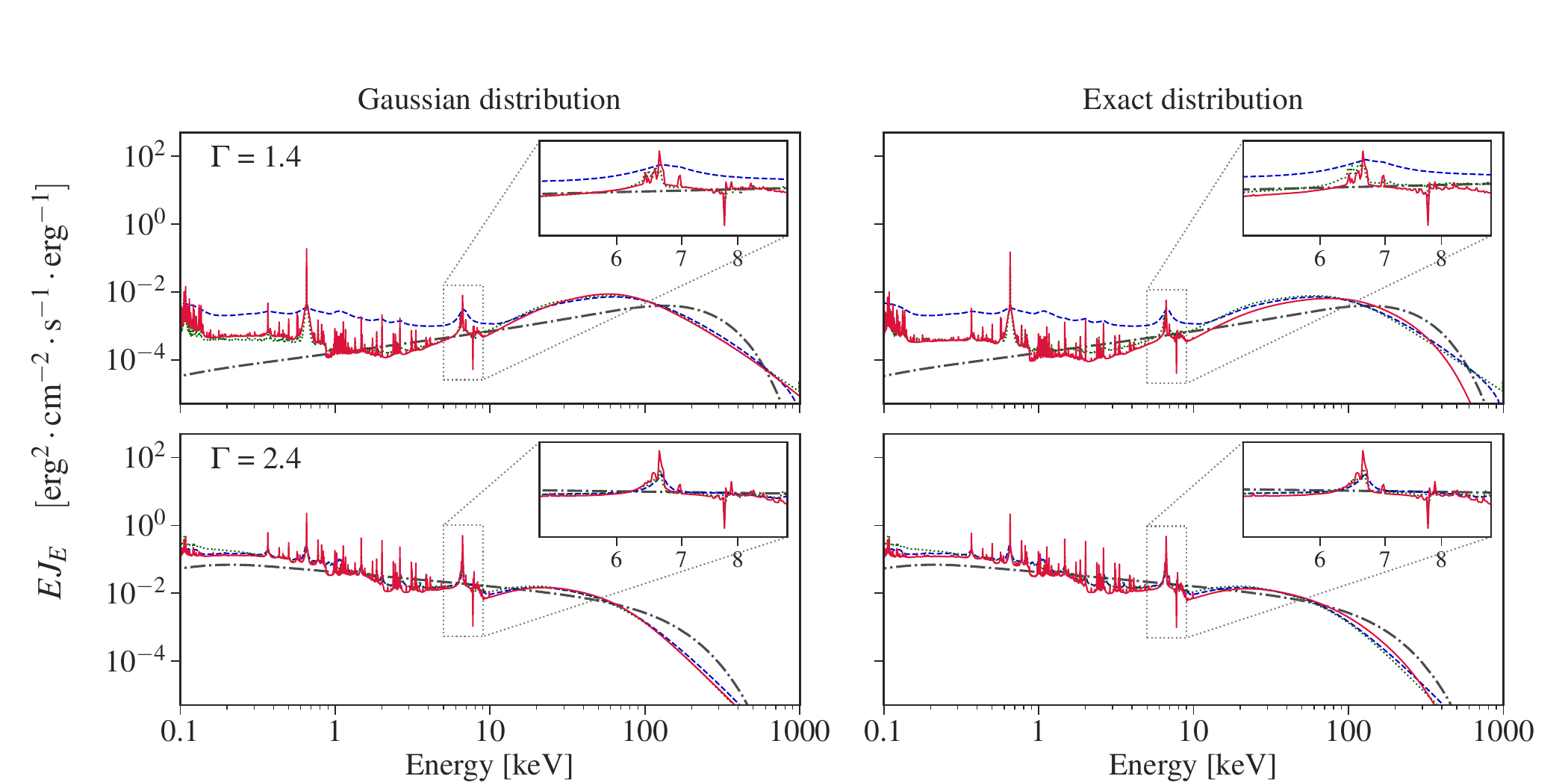}
    \caption{Same as Figure~\ref{Comparsion different model with nthcomp incident}, but with different \texttt{nthcomp} incident spectra obtained from \texttt{xillvercp} by setting \texttt{refl\_frac} = 0, $kT_{\rm e} = 60$~keV, and $\Gamma = 1.4$ (top panel) and $\Gamma = 2.4$ (bottom panel). The ionization parameter is set to $\log\xi = 3.0$ for all spectra.}
    \label{Comparsion different model with nthcomp incident G14}
\end{figure*}

\section{results}\label{sec:result}
\subsection{Comparison with Other Models}\label{Comparison with other model}

First, we employ the \texttt{nthcomp} model \citep{1996MNRAS.283..193Z,1999MNRAS.309..561Z} as the illuminating source at the top boundary. We set $I_\mathrm{type} = \mathrm{file}$ and obtain the incident spectrum by setting the reflection fraction to 0, $\Gamma=2.0$, and $kT_E=60$~keV in \texttt{xillvercp}; this spectrum is then adopted as the standard incident radiation field in this paper. The results are compared with the \texttt{reflionx\_hd\_nthcomp} \citep{2007MNRAS.381.1697R,2020MNRAS.498.3888J} and \texttt{xillvercp} tables. To match these models, we do not adopt any bottom incident radiation field; thus, we set $B_\mathrm{sw} = 0$. Our model and \texttt{xillvercp} adopt the solar abundances table from \cite{1996ASPC...99..117G} for elements from H to Zn, in contrast to the \texttt{reflionx} model, which uses the abundances from \cite{1983ApJ...270..119M}. Accordingly, in this section, we set $A_\mathrm{Fe}$ in our model and \texttt{xillvercp} to 1.32 to match the setting in \texttt{reflionx}. In addition, we set $\log\xi = 1.0$ (top row), 2.0 (middle row), and 3.0 (bottom row), and adopt both the Gaussian approximation (left column) and the exact energy redistribution function (right column) for Compton scattering in Figure~\ref{Comparsion different model with nthcomp incident}. All other parameters are set to their default values listed in Table~\ref{model parameters}.

The spectra in Figure~\ref{Comparsion different model with nthcomp incident} are plotted in the form of $EJ_E$. The \texttt{reflionx} model assumes an isotropic radiation field, whereas both \texttt{DAO} and \texttt{xillver} compute an anisotropic radiation field. For \texttt{DAO} and \texttt{xillver}, the mean intensity is evaluated as $J(E) = \int I(\mu, E)\,d\mu$, where $\mu$ is sampled according to the internal angular discretization adopted in each model\footnote{If we choose $N_\mu=10$, our angle grid is the same as \texttt{xillver}'s: $\mu = \{0.05, 0.15, 0.25, 0.35, 0.45, 0.55, 0.65, 0.75, 0.85, 0.95\}$.}.
As illustrated in Figure~\ref{Comparsion different model with nthcomp incident}, the outputs of the three models are consistent within the 0.1 to 1000 keV range. All three models exhibit a prominent Compton hump, strong absorption edges and emission lines. In particular, in the spectrum generated using the exact redistribution, the Compton hump drops more rapidly at higher energies. In Section~\ref{dif kernel}, we analyze in detail the impact of different redistribution functions. In principle, for the spectra calculated by Gaussian-approximated redistribution function, our results should be identical to xillverCp since we employ the same mathematical formalism, but in practice the different versions of XSTAR employed and other small differences in assumptions are enough to cause small discrepancies in the outputs of the two models. Still, as expected, DAO agrees with xillver better than with reflionx.

In the low-ionization atmosphere ($\log\xi = 1.0$), we observe both the iron K$\alpha$ line and the K$\beta$ line in \texttt{DAO} and \texttt{xillvercp}. The intensities of the iron lines in \texttt{DAO} are very close to those in \texttt{xillvercp} in this case. However, due to limitations in its atomic database, \texttt{reflionx} does not include any iron K$\beta$ transitions. For the intermediate-ionization atmosphere ($\log\xi = 2.0$), the Fe K$\alpha$ emission profile resolves into a complex structure comprising three discrete features in both \texttt{DAO} and \texttt{xillvercp}: a dominant peak at 6.43 keV (corresponding to low-ionization $\mathrm{Fe}_\mathrm{VII–XVII}$), a secondary component near 6.51 keV (attributable to intermediate L-shell ions, $\mathrm{Fe}_\mathrm{XVIII–XX}$), and a distinct high-energy peak at 6.58 keV (consistent with $\mathrm{Fe}_\mathrm{XXI–XXII}$). This profile reveals that the plasma at $\log\xi = 2.0$ comprises a rich mixture of various iron ionization states. For the highly ionized atmosphere ($\log\xi = 3.0$), we observe the $\mathrm{Fe}_\mathrm{XXV}$ and $\mathrm{Fe}_\mathrm{XXVI}$ $\mathrm{K}\alpha$ lines, as well as a high-energy feature around 7.9 keV. This latter feature is identified as a blend of the $\mathrm{Ni}_\mathrm{XXVII}$ $\mathrm{K}\alpha$ line and the $\mathrm{Fe}_\mathrm{XXV}$ $\mathrm{K}\beta$ transition. Figure \ref{fig:ni} reveals a significant coexistence of helium-like Fe and Ni ions within the highly ionized atmosphere. Regarding the absence of Nickel features at lower ionization states: The high-energy feature observed at $\log\xi = 3.0$ cannot be observed in intermediate- or low-ionization atmospheres. In these cases, the Nickel line energy consistently falls above the Iron K-edge energy. Since Nickel is much less abundant than Iron ($\sim$1/17), the strong absorption from Iron suppresses the Nickel emission, making the line invisible. In contrast, this $\mathrm{Ni}_\mathrm{XXVII}$ $\mathrm{K}\alpha$ feature is absent in the \texttt{xillvercp} and \texttt{reflionx} table models, as these models do not include atomic calculations for Nickel. Our results in high ionization parameter are consistent with \cite{2024ApJ...974..280D}, which take Nickel into account in \texttt{xillver} model.

The detectability of Nickel in astrophysical spectra is well-supported by observations. Narrow Ni K$\alpha$ emission lines have been robustly detected in the X-ray spectra of bright AGNs \citep{2003MNRAS.343L...1M}. In the regime of High-Mass X-ray Binaries, detailed spectroscopy of GX 301-2 using Suzaku \citep{2012ApJ...745..124S} has explicitly resolved the $\mathrm{Ni}$ $\mathrm{K}\alpha$ line at $\approx 7.47$ keV, separating it from the dominant iron complex. Consequently, in highly ionized environments where the Fe K-edge absorption is diminished, the $\mathrm{Ni}$ $\mathrm{K}\alpha$ line predicted by our model should be observable, particularly with the high-resolution capabilities of future missions like \textit{NewAthena}.

\begin{table*}
    \centering
    \begin{tabular}{c|c|c}
    \hline
    \hline
         Model& Atomic data table& {\tt XSTAR} version \\
         \hline    
         \hline
             
         \texttt{DAO}& ATDB (2024-10-17T14:09:13)\citep{2021Atoms...9...12M}&2.59\citep{2021ApJ...908...94K}\\
  
         \texttt{xillvercp} & ATDB (2012-08-03T01:58:54) & 2.2.1bn \citep{2001ApJS..133..221K} \\

         \texttt{reflionx}   & custom definition \citep[see][]{2005MNRAS.358..211R} & not used\\
         \hline
    \end{tabular}
    \caption{Atomic physics treatments in the three models}
    \label{Atom table version}
\end{table*}
 \begin{figure*}
    \centering
    \includegraphics[width=\linewidth]{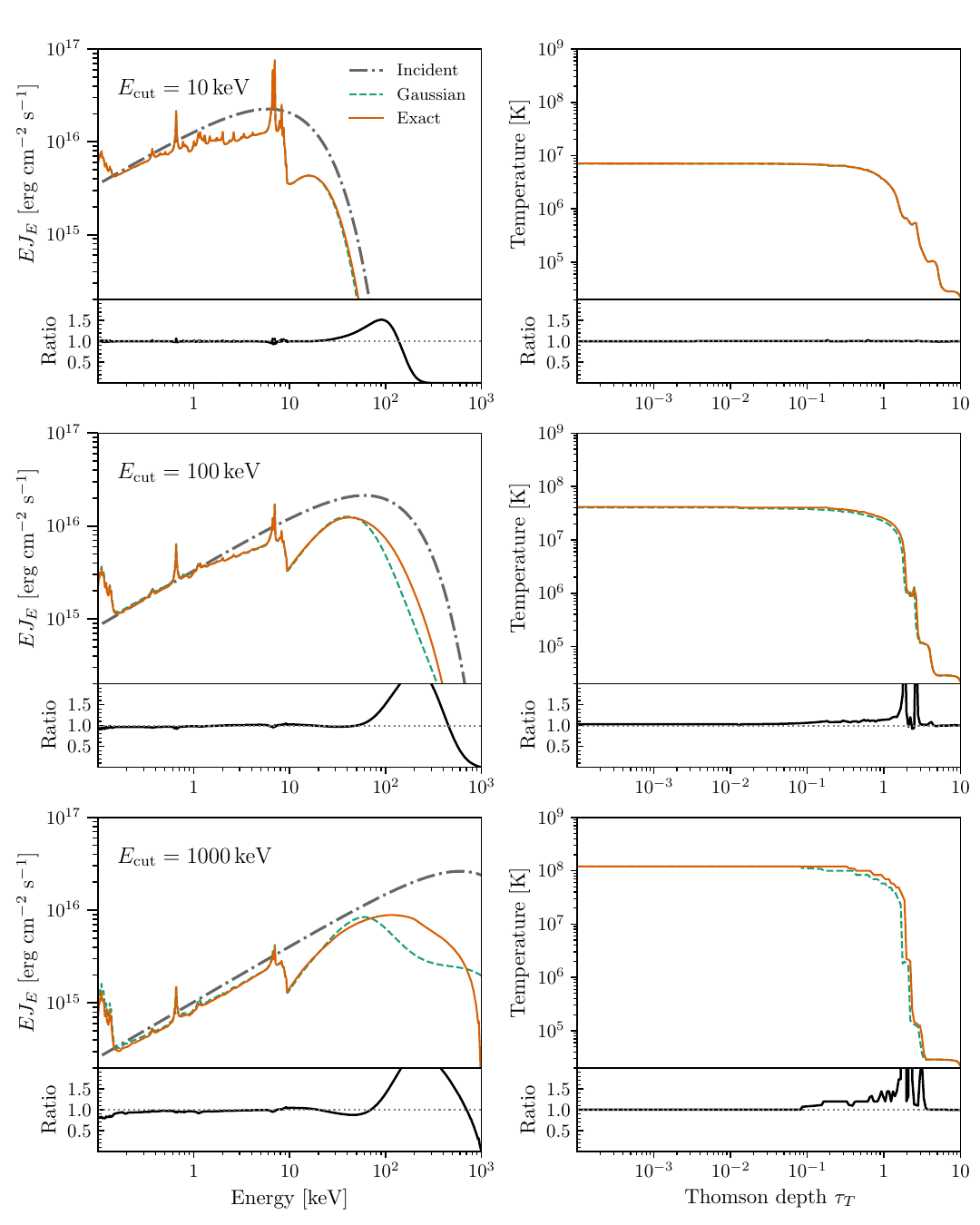}
    \caption{Comparison of angle-averaged reflection spectra computed using the exact (red solid line) and Gaussian (green dashed line) redistribution functions. The incident cut-off power-law spectrum is shown as a grey dashed line. The parameters are $\Gamma = 1.4$, $A_{\mathrm{Fe}} = 5.0$, and $\log\xi = 3.5$, with varying cut-off energies of $E_\mathrm{cut} = 10$~keV (top), 100~keV (middle), and 1000~keV (bottom). Other parameters are set at their default value. The lower sub-panel in each plot displays the ratio of the spectrum computed with the exact redistribution to that computed with the Gaussian approximation.}
    \label{fig:different kernel}
\end{figure*}
We also test our model with a softer and a harder illumination, specifically $\Gamma=1.4$ and $\Gamma=2.4$, with the results presented in Figure~\ref{Comparsion different model with nthcomp incident G14}. In Figure~\ref{Comparsion different model with nthcomp incident G14}, the ionization parameter is set at 3.0 for all spectra; the other parameters are the same as in Figure~\ref{Comparsion different model with nthcomp incident}. Hard X-ray illumination with exact scattering redistribution displays a notable difference in the Compton hump from that of the Gaussian-approximated redistribution. This is caused by the inaccuracy of the Gaussian-approximation redistribution function for high-energy photons and electrons (see Section~\ref{dif kernel} for more details). However, we also find noticeable discrepancies among the three models in the soft X-ray band (0.1--1 keV) for any \texttt{nthcomp} incident spectrum. Light-element $\alpha$ lines (e.g., C, N, O) as well as Fe L-shell emission lines are generally present in this energy range. For \texttt{reflionx}, such discrepancies are expected because the model includes only a limited set of ions. In contrast, the differences in the soft X-ray range between \texttt{xillver} and \texttt{DAO} most likely arise from the use of different atomic databases and different versions of {\tt XSTAR} in their calculations. Table~\ref{Atom table version} summarizes the {\tt XSTAR} versions and atomic data tables employed by each code. The \texttt{ATDB 2024} release incorporates corrections for high-density plasma effects and improved radiative and collisional rates for odd-$Z$ elements, with related developments reported in \citet{2017A&A...604A..63M,2018A&A...616A..62M,2021Atoms...9...12M}. Moreover, \citet{2024ApJ...974..280D} demonstrated the impact of the latest {\tt XSTAR} and \texttt{ATDB} versions on reflection spectra, as shown in their Figures 3 and 4. They also found that using different versions of \texttt{XSTAR} and atomic databases in the \texttt{xillver} model produced noticeable differences.

In summary, \texttt{DAO} incorporates the more accurate treatment of atomic processes originally explored in \cite{2024ApJ...974..280D} by employing the latest versions of {\tt XSTAR}, and the precise description of Compton scattering first explored in \cite{2020ApJ...897...67G} into a publicly available code that can be run in bespoke configurations. Furthermore, we compare \texttt{DAO} with \texttt{pexrav} \citep{1995MNRAS.273..837M} to validate our treatment of Compton scattering. The results are shown in Figure \ref{fig:pexrav} in Appendix \ref{apd:algorithm}. Although the \texttt{pexrav} spectrum exhibits a slightly higher peak in the Compton hump, the two models are consistent at energies $E < 20$ keV and $E > 50$ keV.

\subsection{Effects of Energy Redistribution Functions}\label{dif kernel}

Compton scattering is a stochastic process that alters both the propagation direction and the energy of photons, and it should be described using the exact quantum mechanical formalism. 


However, all publicly available reflection table models adopt a Gaussian approximation \citep{1978ApJ...219..292R} to treat this process, which is only valid to describe the energy redistribution for low-energy photons ($E_{\rm ph} \ll m_e c^2$) scattered by low-temperature electron clouds ($kT_e \ll E_{\rm ph}$). Appendix Figure~\ref{fig:kernel} illustrates the differences between the exact and Gaussian redistribution functions. (1) At low electron temperatures and for soft incident photons ($E_i = 1$~keV), both the Gaussian and exact redistribution functions exhibit a single peak with good symmetry. (2) When the incident photon energy approaches $m_e c^2$ while the electron temperature remains low, the exact redistribution function develops a double-peaked and asymmetric structure. (3) At sufficiently high electron temperatures (e.g., $T = 10^9$~K), the Gaussian approximation becomes excessively broadened and no longer captures the actual shape of the Compton scattering redistribution. \citet{2020ApJ...897...67G} investigated this effect on the reflection spectrum and included it into the proprietary \texttt{xillver} source code, but it has not yet been implemented in a publicly released table model. \texttt{DAO} incorporates the publicly available source code released by \citet{2020ApJ...897...67G} to include the exact Compton scattering kernel. 

The exact redistribution function for Compton scattering has been extensively studied. \citet{1981MNRAS.197..451G} incorporated the relativistic Maxwellian velocity distribution into the redistribution function, but a computational error was present in the original paper. Subsequently, \citet{1993AstL...19..262N}, \citet{1996ApJ...470..249P}, and \citet{2010ApJS..189..286P} investigated the redistribution function for photons scattered by anisotropic electrons using quantum electrodynamical methods. Later, \citet{2017MNRAS.469.2032M} corrected the earlier error in \citet{1981MNRAS.197..451G} and demonstrated consistency among these methods. \citet{2020ApJ...897...67G} used the method described in \citet{1993AstL...19..262N} and showed reflection spectra using the exact redistribution function. We present the algorithms for the Gaussian-approximated and quantum-mechanical redistribution functions in Appendix~\ref{apd:redistribution function}.

Figures~\ref{Comparsion different model with nthcomp incident}--\ref{Comparsion different model with nthcomp incident G14} display that a discrepancy exists between the reflection spectra generated by the exact redistribution and those from the Gaussian approximation within the hard X-ray band. We investigate this further in Figure~\ref{fig:different kernel}. For the purposes of comparison with \texttt{xillver}, we adopt the same assumptions as Fig 7 of \citet{2020ApJ...897...67G}, in which the exact redistribution function was investigated within the \texttt{xillver} model. Specifically, we consider \texttt{DAO} with a cut-off power-law as the incident spectrum with $\Gamma = 1.4$ and a variety of cut-off energies ($E_{\rm c}=10$, 100, 1000~keV). To show how this could impact the Compton shoulder of K$\alpha$ lines at 6.4~keV and Compton hump, the iron abundance is set to 5.0 and the ionization parameter to 3.5. Other parameters are set at default values. The angle-averaged spectra shown in the left column and the temperature profiles in the right column clearly illustrate how these differences depend on temperature. The surface temperature is highest for the hardest incident spectrum ($E_{\rm c} = 1000~\mathrm{keV}$), and in this case the differences in the reflection spectrum between the two methods are most pronounced. However, for the coolest gas, where the surface temperature is an order of magnitude lower than in the former case, the impact of the inaccurate Gaussian redistribution function is smaller, but can still be observed in the ratio plot. Compared with the large difference in the Compton hump region, the deviation of the Compton shoulder is not very obvious for all three models. \cite{2020ApJ...897...67G} also presented reflection spectra assuming cut-off power-law illumination in their Fig 7. The temperature profile calculated by \texttt{DAO} (left column in Figure \ref{fig:different kernel}) is significantly different from theirs, being approximately an order of magnitude lower. In the case of power-law or cut-off power-law illumination, the \texttt{xillver} model imposes an artificial low-energy cut-off, which results in a harder illuminating spectrum. In contrast, \texttt{DAO} does not currently apply any low-energy cut-off.

\begin{figure*}
    \centering
    \includegraphics[width=\linewidth]{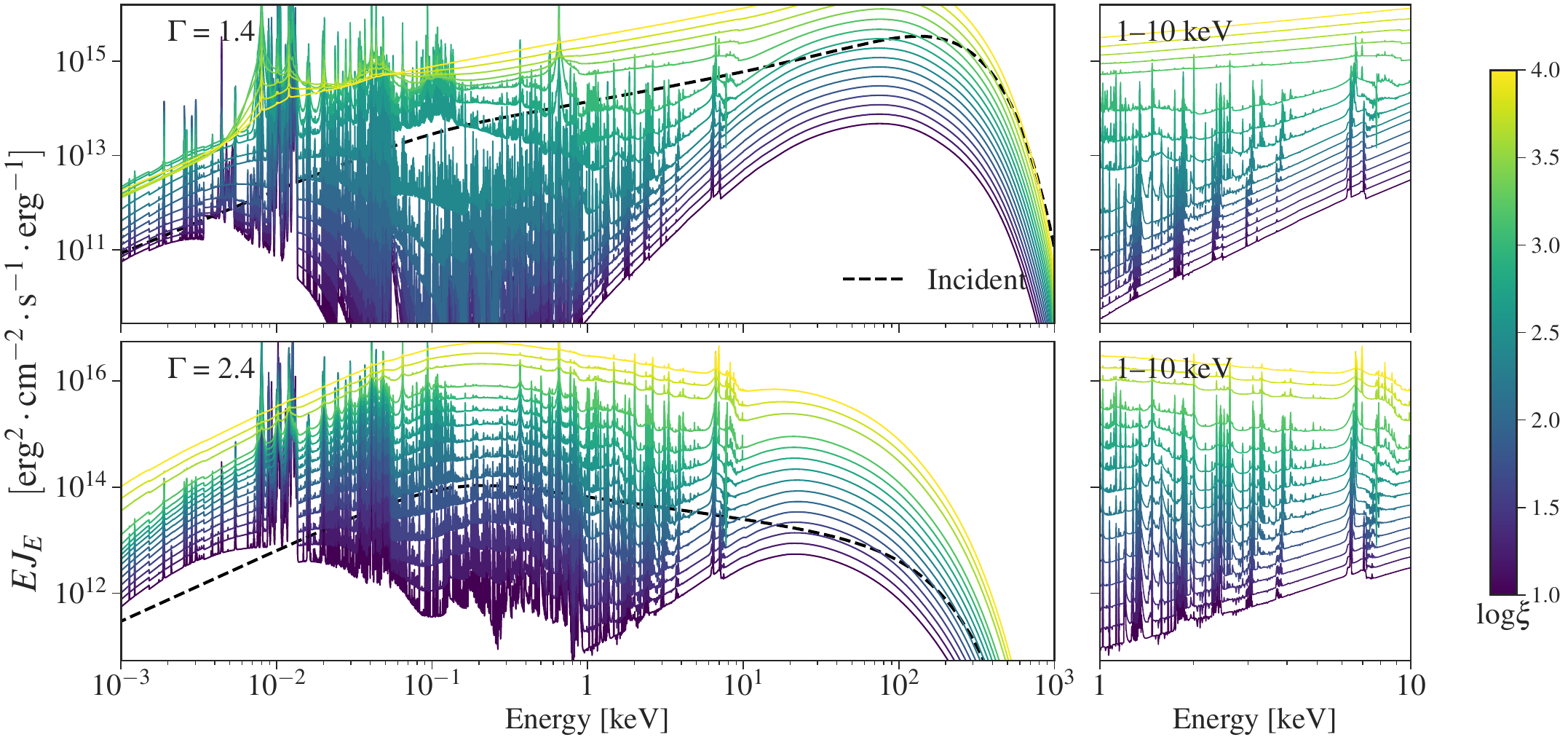}
    \caption{Angle-averaged reflection spectra for various ionization parameters ($\log\xi$), distinguished by color. The \texttt{nthcomp} incident spectra are obtained from \texttt{xillvercp} by setting \texttt{refl\_frac} = 0 and $kT_E = 60$~keV, with $\Gamma = 1.4$ (top panel) and $\Gamma = 2.4$ (bottom panel). Other parameters are set at their default values. The shape of incident spectrum is shown as a black dashed line for comparison.}
    \label{fig:different ionization parameter}
\end{figure*}

\begin{figure}[h!]
    \centering
    \includegraphics[width=\linewidth]{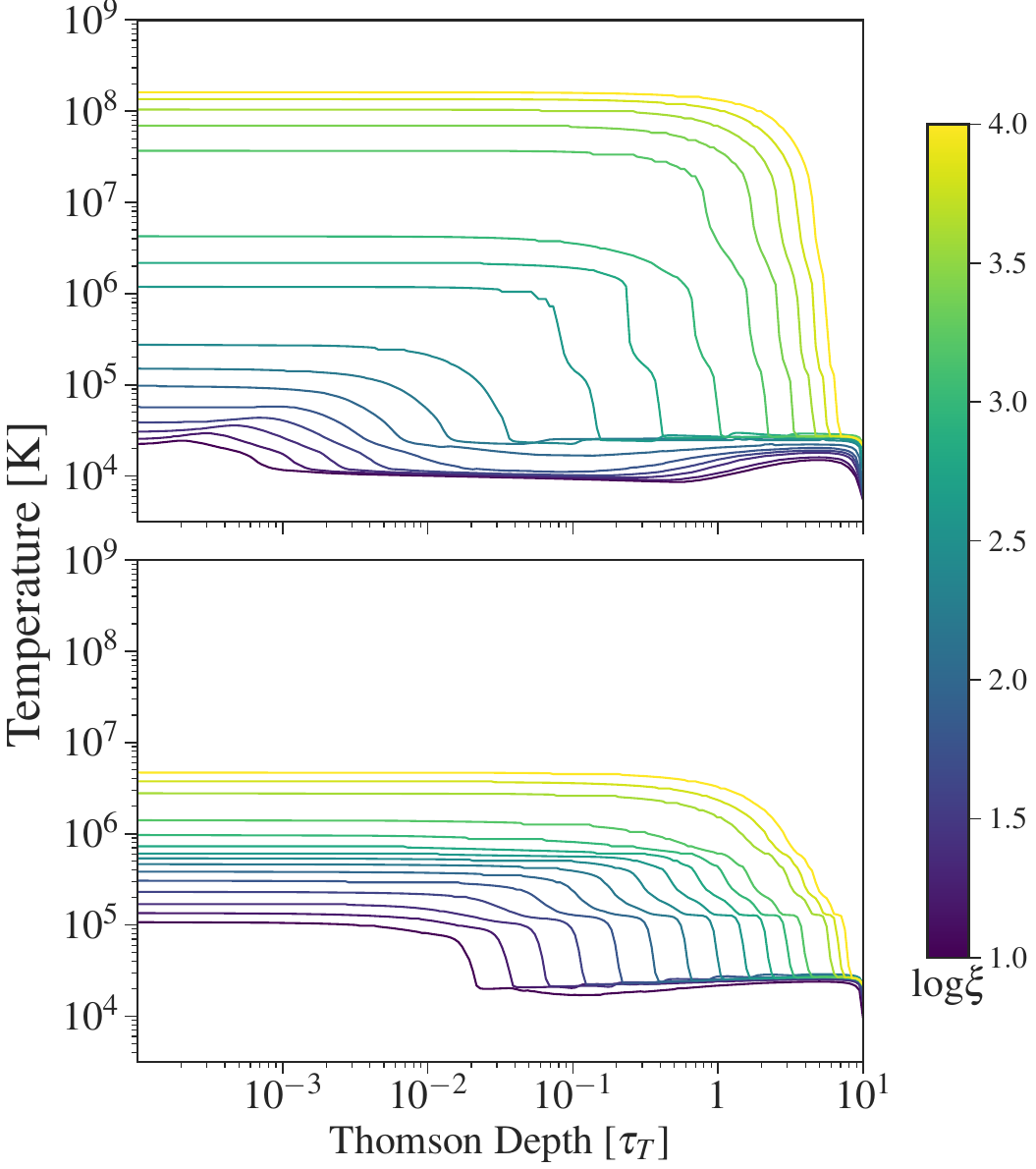}
    \caption{Gas temperature for different ionization parameters, parameters set as same as Figure \ref{fig:different ionization parameter}.}
    \label{fig:temperature for different ionization parameter}
\end{figure}

Current instruments, such as \textit{NuSTAR}, can image the sky in the 3--79~keV band, while \textit{Insight}-HXMT extends observations to hard X-ray energies up to 250~keV \citep{2020SCPMA..6349502Z}. Both of them can observe the hard X-ray band, where deviations in measurements arise, especially for hard spectra and high coronal temperatures \citep{2017MNRAS.468.3489K,2019MNRAS.490.1350B}. We therefore employ the exact redistribution function for all subsequent calculations.

\subsection{Effects of varying the ionization parameter}\label{chap:ionization}
\begin{figure*}
    \centering
    \includegraphics[width=\linewidth]{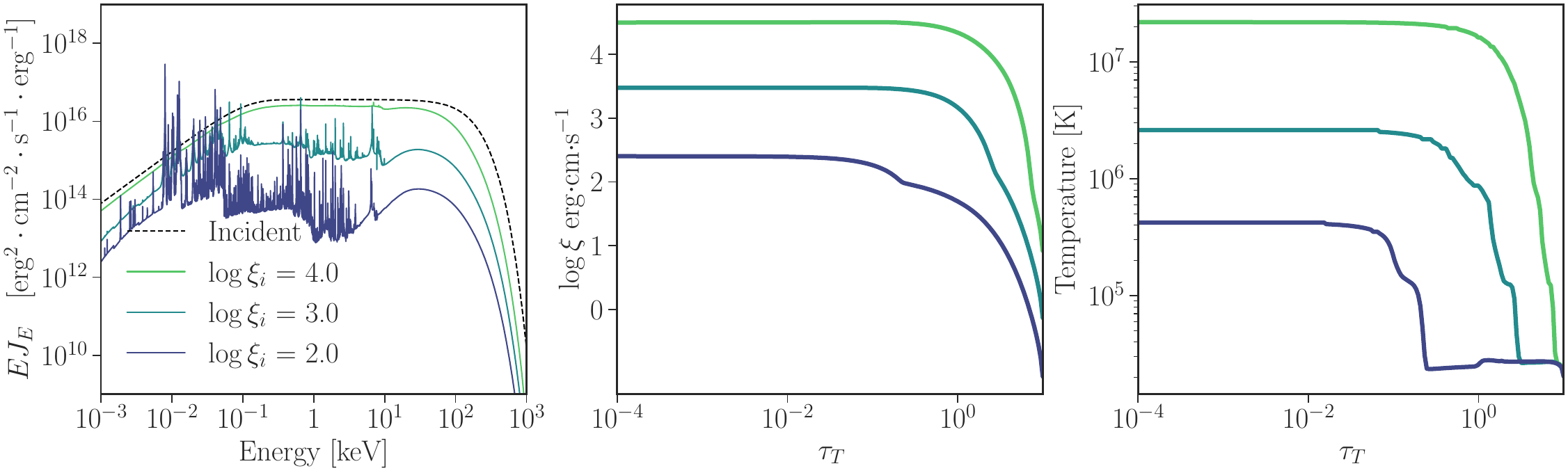}
    \caption{Left: Angle-averaged reflection spectra for different ionization parameters ($\log\xi_i$). The incident radiation is a Comptonization spectrum with $\Gamma = 2.0$. The incident spectrum for the $\log\xi_i = 4.0$ case is plotted in the left panel for reference. Middle and Right: Profiles of the ionization parameter (Middle) and gas temperature (Right) as a function of optical depth.}
    \label{fig:xivstau}
\end{figure*}
\begin{figure*}
    \centering
    \includegraphics[width=\linewidth]{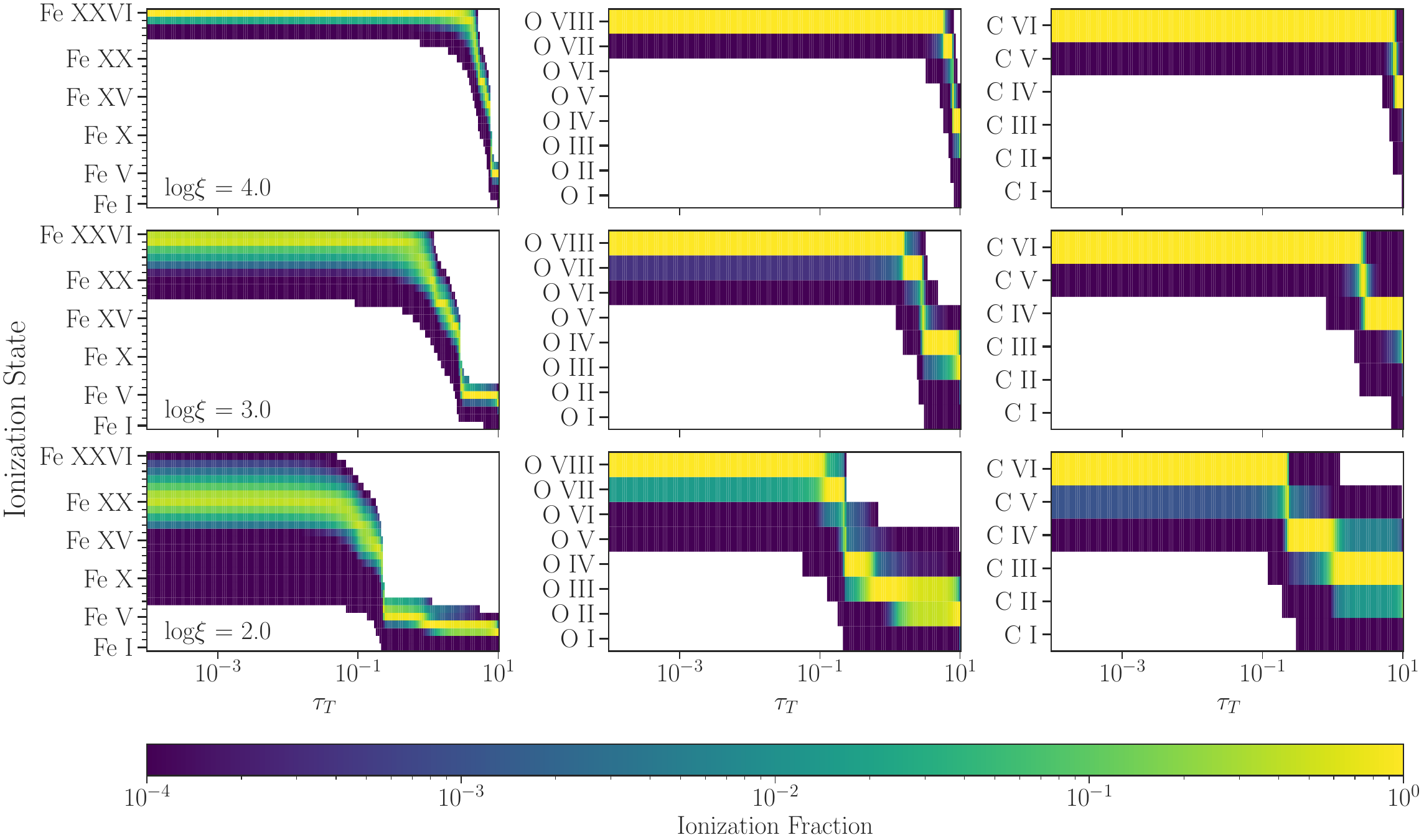}
    \caption{Ionic fractions of iron (Fe), oxygen (O), and carbon (C) as a function of Thomson optical depth for three different ionization parameters: $\log\xi = 4.0$ (top row), $\log\xi = 3.0$ (middle row), and $\log\xi = 2.0$ (bottom row). The other parameters are the same as in Figure~\ref{fig:xivstau}.}
    \label{fig:elements fraction}
\end{figure*}
We assume a constant-density atmosphere, and the form of \cite{1969ApJ...156..943T} is used to calculate the ionization parameter $\xi$ (Eq.~\ref{eq:ionization par}). At each optical depth layer, the total ionizing flux $F_x$ of the radiation field is calculated to obtain the ionization parameter. Figure~\ref{fig:different ionization parameter} illustrates the emergent spectra for various ionization parameters, considering the exact redistribution function for Compton scattering. No normalization has been applied to the final spectra displayed in this plot. It is evident that spectra corresponding to a higher ionization parameter exhibit greater luminosity, whereas spectra with a lower ionization parameter display more prominent atomic line features and the Compton shoulder.

Figure \ref{fig:temperature for different ionization parameter} shows the temperature profile as a function of the Thomson optical depth (from 10$^{-4}$ to 10) for the two incident spectra shown in Figure \ref{fig:different ionization parameter}. In a highly ionized atmosphere, the dominant heating process is electron recoil following Compton scattering. The net heating rate is given by \cite{1978ApJ...219..292R}
\begin{equation}\label{eq:Compton heating}
    n_e\Gamma_e = \frac{\sigma_T}{m_ec^2}\left(\int_0^\infty EJ_E dE - 4kT\int J_E dE\right)
\end{equation}
where $\Gamma_e$ is the net Compton heating rate, $n_e$ is the electron density, $E$ is the energy, $J_E$ is the mean intensity, $\sigma_T$ is the Thomson cross section, $m_ec^2$ is the electron rest energy, $k$ is the Boltzmann constant, and $T$ is the electron temperature. The Compton heating rate is related to $EJ_E$, which implies a large heating rate when there are more photons in the hard X-ray band. In a less ionized atmosphere, the dominant heating process is photoionization, which is sensitive to the soft X-ray band. Consequently, the surface temperature for the lowest ionization parameter ($\log\xi = 1.0$) under a soft X-ray incident spectrum is higher than that under a hard X-ray incident spectrum. Conversely, the surface temperature for the highest ionization parameter ($\log\xi = 4.0$) under a hard X-ray incident spectrum is higher than that under a soft X-ray incident spectrum because of Compton heating.


\begin{figure}
    \centering
    \includegraphics[width=\linewidth]{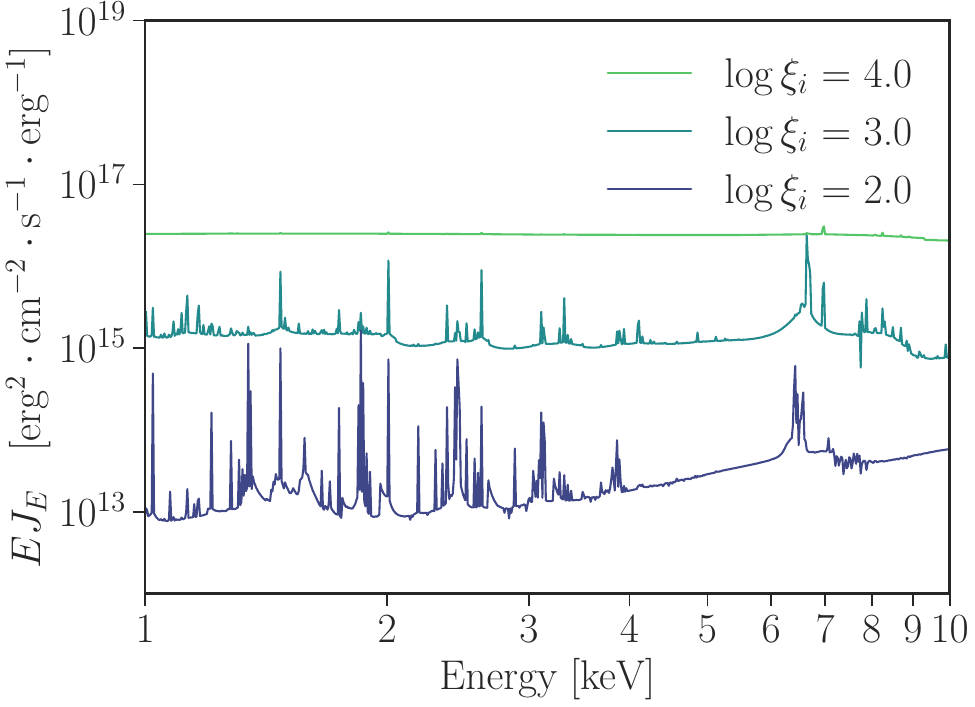}
    \caption{Evolution of the reflection spectrum in the 1--10~keV band as a function of the ionization parameter $\xi$. The color gradient represents the transition from low ionization (dark blue, $\log\xi=2.0$) to a nearly fully ionized plasma (bright green, $\log\xi=4.0$). The lower ionization models exhibit a rich spectrum of K$\alpha$ and K$\beta$ emission lines, whereas these features are significantly diminished by the strong Compton scattering continuum in the highest ionization case.}
    \label{fig:iron line}
\end{figure}

In summary, the ionization parameter affects how photons interact with electrons and ions in the gas. Such effects change the temperature profile and the spectral shape. In Figure~\ref{fig:xivstau}, we show the reflection spectra (left), the ionization parameter (middle), and the temperature profile (right) as a function of optical depth in the final iteration step. Corresponding ionic fractions for iron (Fe), oxygen (O), and carbon (C) are visualized as colormaps in Figure~\ref{fig:elements fraction}. The incident spectrum for these plots is a Comptonization spectrum with $\Gamma = 2.0$ as shown in the first panel of Figure~\ref{fig:xivstau}.

Figure~\ref{fig:elements fraction} shows that multiple intermediate and highly ionized states of Fe, C, and O appear in the upper layers ($\tau_T < 0.1$) of a weakly ionized disk ($\log \xi = 2.0$). As the depth increases, these ions gradually transition to lower ionization states due to the decrease in the ionizing flux. The high abundance of these ions in the upper layers leads to strong K-, L-, and M-shell transition lines in the emergent reflection spectra (see Figure~\ref{fig:xivstau}). In contrast, at the highest ionization parameter ($\log \xi = 4.0$), the ionization balance shifts strongly toward more highly stripped ions, driving the transition zones of each element to greater depths within the slab. For example, in such a highly ionized disk, only H-like and He-like species such as Fe~\textsc{xxvi} and Fe~\textsc{xxv} persist (Figure~\ref{fig:elements fraction}, top row). As a result, the K$\alpha$ line at 6.4~keV from neutral or low-ionization iron is entirely absent in the reflection spectrum (Figure~\ref{fig:xivstau}, bright green line). Furthermore, the equivalent widths of the remaining K$\alpha$ lines from Fe~\textsc{xxvi} ($\sim$6.97~keV) and Fe~\textsc{xxv} ($\sim$6.7~keV) are significantly diminished by the strong Compton scattering continuum (see Figure~\ref{fig:iron line} for details). In contrast, for the lower ionization cases of $\log\xi = 2.0$ and 3.0, the 1--10~keV band exhibits a much richer emission-line spectrum. This energy range is populated not only by the complex iron K$\alpha$ and K$\beta$ transition series but also by K-shell transitions from lighter elements (e.g., S, Ar, Ca) as a wider range of ionization states contribute. Our results here are consistent with similar analyses of other reflection models \citep{2005MNRAS.358..211R,2010ApJ...718..695G,2013ApJ...768..146G}.

\begin{figure*}[htbp]
    \centering
    \includegraphics[width=\linewidth]{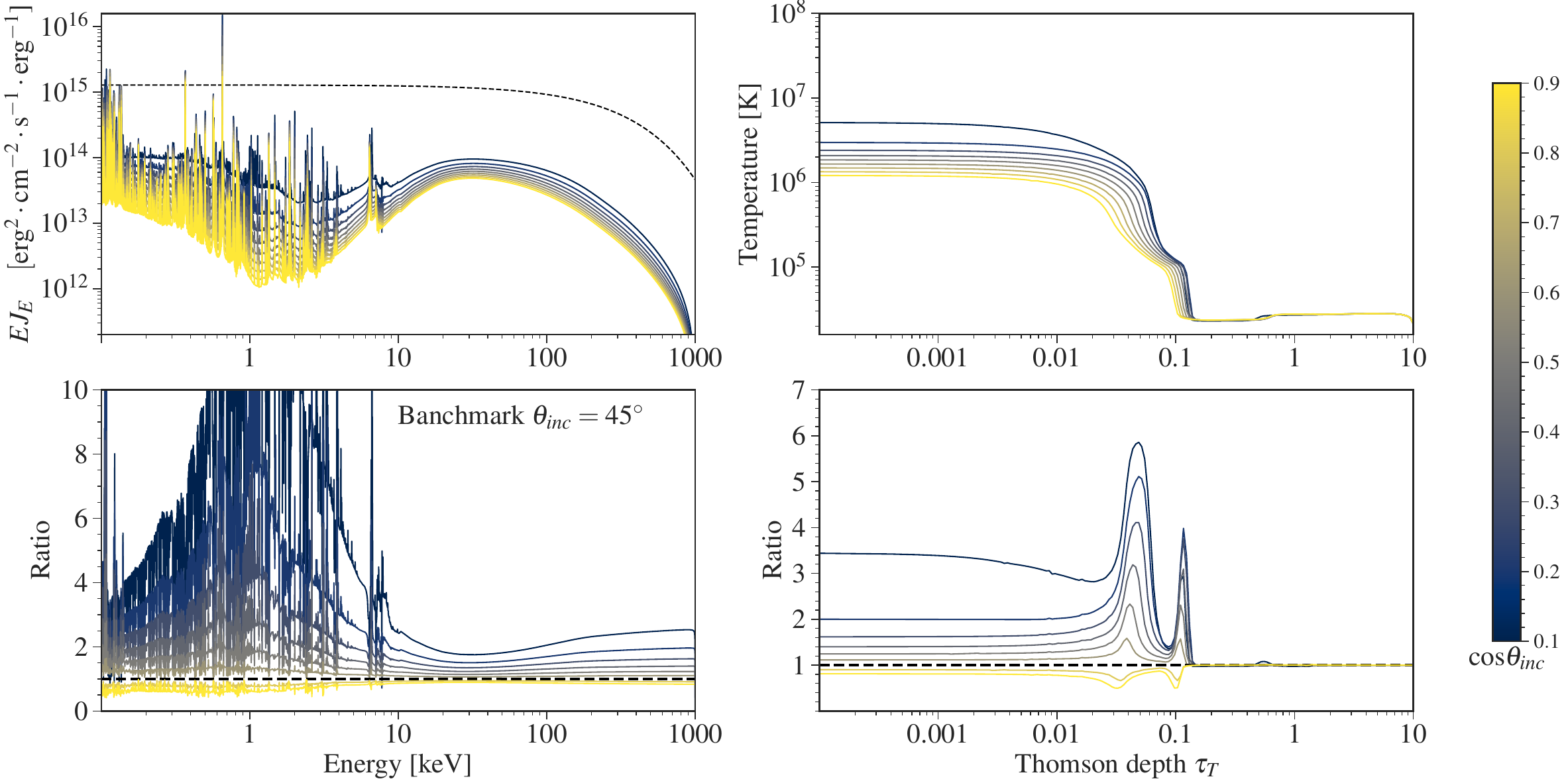}
    \caption{Top row: Angle-averaged emergent intensity (left) and the corresponding temperature profiles (right), color-coded by the incidence angle $\cos\theta_\mathrm{inc}$. Bottom row: Spectral ratios relative to the $\theta_{\mathrm{inc}} = 45^\circ$ case for all incidence angles. The incidence angles are sampled linearly in $\mu = \cos\theta_{\mathrm{inc}}$ between 0 and 1 using 30 bins. The ionization parameter is $\log\xi = 2.0$, and the incident radiation field (black dashed line) is a cut-off power-law with $\Gamma = 2.0$ and $E_\mathrm{cut} = 300~\mathrm{keV}$. The other parameters are set at their default values.}
    \label{fig:different incident angle}
\end{figure*}

\subsection{Effects of varying the incident angle}\label{dif incidentang}

In our model, following the \texttt{xillver} models, the coronal illumination field is specified by a single incidence angle. The $I_\mathrm{inc}$ at the top boundary of the second-order radiative transfer equation (Eq.~\ref{eq:boundary}) is set to zero everywhere except at $\mu_{\mathrm{inc}}$. Since the flux is defined as the first moment of the intensity,
\begin{equation}
    F_x(\tau,E) = \int_0^1 u(\tau,E,\mu) \mu d\mu,
\end{equation}
where $\mu = \cos\theta$, and we adopt
\begin{equation}
    I = I_\mathrm{inc}\delta(\mu-\mu_\mathrm{inc}).
\end{equation}
The incident intensity at the top boundary can then be evaluated as
\begin{equation} \label{eq:incident intensity}
    I_{\mathrm{inc}}(E) = \frac{2F_x(E)}{\mu_{\mathrm{inc}}}.
\end{equation}
In \texttt{DAO}, we always calculate $F_x(E)$ first and then use Eq.~\ref{eq:incident intensity} to derive $I_{\mathrm{inc}}$ for the boundary condition (Eq.~\ref{eq:boundary}). This relation shows that the incidence angle $\mu_{\mathrm{inc}}$ directly affects the illumination intensity, which in turn influences the radiative transfer solution. Consequently, a smaller $\mu_{\mathrm{inc}}$ is expected to produce higher gas temperatures and a higher degree of ionization at the surface.

Figure~\ref{fig:different incident angle} shows the emission-angle-averaged flux for different incidence angles. Incidence angles are sampled from 0.1 to 0.9 in steps of 0.1 (color-coded from purple to yellow). The illumination radiation field is modeled with a cut-off power-law, using $\Gamma = 2.0$ and $E_{\mathrm{cut}} = 300~\mathrm{keV}$. The ionization parameter is set to 2.0, and the number of angular grids is fixed at $N_{\mu} = 30$ to improve the angular resolution, while all other model parameters are set to their default values. The results are broadly consistent with our expectation and with previous studies of the \texttt{xillver} model \citep{2010ApJ...718..695G,2013MNRAS.430.1694D}: the gas temperature and the strength of the radiation hump show a clear inverse correlation with $\cos\theta_{\mathrm{inc}}$.

On the other hand, the incidence angle directly affects the radiative transfer solution. To focus on this effect, we consider a purely scattering atmosphere with a fixed gas temperature of $10^9$~K, thereby suppressing variations in ionization and temperature profile that would otherwise arise from changes in the effective radiation intensity. This setup allows us to focus solely on the role of the incidence angle. In this case, the source function can be expressed as
\begin{equation}
    S(E) = \frac{1}{2\sigma(E)} \int dE' R(E',E) \int_{-1}^{1}d\mu I(E',\mu)
\end{equation}
which is the second term in Eq. \ref{eq:source_function} with $\alpha(E) = 0$.

\begin{figure}[htbp]
    \centering
    \includegraphics[width=\linewidth]{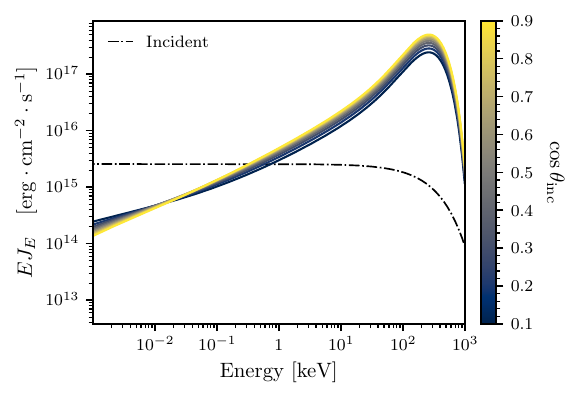}
    \caption{Solution of the radiative transfer equation for a pure scattering atmosphere. The incident spectrum is a cut-off power-law ($\Gamma$ = 2.0, $E_\mathrm{cut} = 300$ keV), and the gas temperature is fixed at $10^9$~K.}
    \label{fig:pure scattering}
\end{figure}

Figure~\ref{fig:pure scattering} displays the angle-averaged emergent spectra. After a few iterations, all spectra reveal a prominent Compton scattering hump at high energies. This hump is weakest for the most grazing incidence angle ($\mu_{\mathrm{inc}} = 0.1$), despite this case having the strongest effective incident radiation (Eq.~\ref{eq:incident intensity}). This behavior is consistent with the reflection spectra in Figure~\ref{fig:different incident angle}, where the intensity of the Compton hump relative to the iron line is significantly greater at small $\mu_\mathrm{inc}$ than at large ones.

In summary, the incidence angle fundamentally influences the reflection spectrum through two distinct mechanisms: 1) it determines the effective radiation intensity incident on the disk, thereby governing the ionization balance and temperature profile, and 2) it is an explicit parameter in the angular dependence of the radiative transfer equation. While the \texttt{xillver} table model neglects the latter, our approach explicitly incorporates this dependence. Notwithstanding this advancement, a key approximation remains in our model: the source function does not yet account for the angular distribution of scattering. The effects of this simplification on data analysis are deferred to a future publication.

\subsection{Emergent intensity along all angle bins}

Depending on the choice of angular binning, the emergent intensity can be obtained for different emission angles \citep[as was first extensively investigated by][] {2014ApJ...782...76G}. In the special case where $\mu_{\mathrm{inc}} = \mu_{\mathrm{emis}}$, the emergent radiation field is given by
\begin{equation}
    I^+({\mu_{\mathrm{emis}}}) = 2 \, u({\mu_{\mathrm{emis}}}) - I^-
\end{equation}
where $I^-$ denotes the illuminating spectrum from the corona. At other emission angles, the emergent radiation field is given by
\begin{equation}
    I^+ _{\mu_\mathrm{emis}} = 2 u_{\mu_\mathrm{emis}}
\end{equation}

\begin{figure}[htbp]
    \centering
    \includegraphics[width=\linewidth]{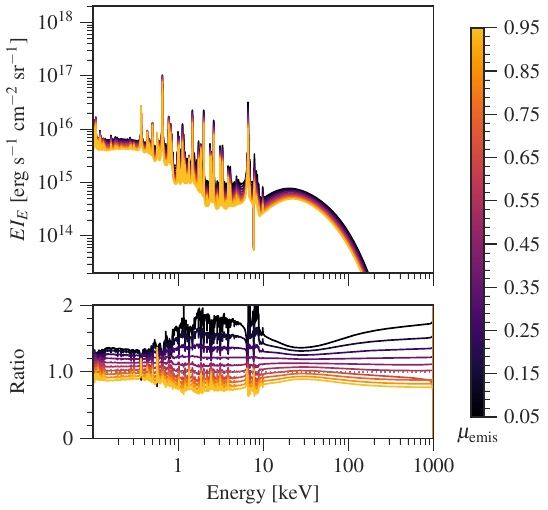}
    \caption{Emergent intensity spectra (top panel) and the ratio of emergent intensity to the angle-averaged intensity (bottom panel) for a fixed incident angle of $\theta_{\mathrm{inc}} = 45^{\circ}$. The profiles are color-coded by the cosine of the emission angle, $\mu_{\mathrm{emis}}$.}
    \label{fig:intensity}
\end{figure}

As shown in Figure~\ref{fig:intensity}, the radiation intensity decreases as $\mu$ increases \citep[in agreement with Fig 2 of][]{2014ApJ...782...76G} This behavior is primarily driven by the effective optical depth, defined as
\begin{equation}
    \tau_{\mathrm{eff}} = \frac{\tau}{\mu}
\end{equation}
Photons escaping at grazing angles (low $\mu$) traverse a longer effective path through the medium. Consequently, spectral features originating from deeper layers, such as emission lines from heavy elements (e.g., Fe), suffer stronger attenuation compared to those produced near the surface (e.g., C, O). For hard X-ray photons, the extended path length increases the probability of scattering, rendering the Compton hump highly sensitive to minor variations in optical depth. This effect, well established in previous works \citep{1981ApJ...248..738L,1988ApJ...335...57L,1995MNRAS.273..837M}, is critical for interpreting the continuum. 

The angular dependence of the emissivity plays a vital role in determining the observed spectrum \citep[e.g.,][]{2014ApJ...782...76G,2025MNRAS.536.2594L}. Recently, \citet{2025ApJ...989..168H} highlighted the impact of emission angles on relativistic spectra shaped by strong gravity. In Appendix~\ref{apd:rel}, we present relativistic spectra calculated using the formalism adopted from the \texttt{relconv}, \texttt{relxill} and \texttt{relxillA} model.

\begin{figure*}[htbp]
    \centering
    \includegraphics[width=\linewidth]{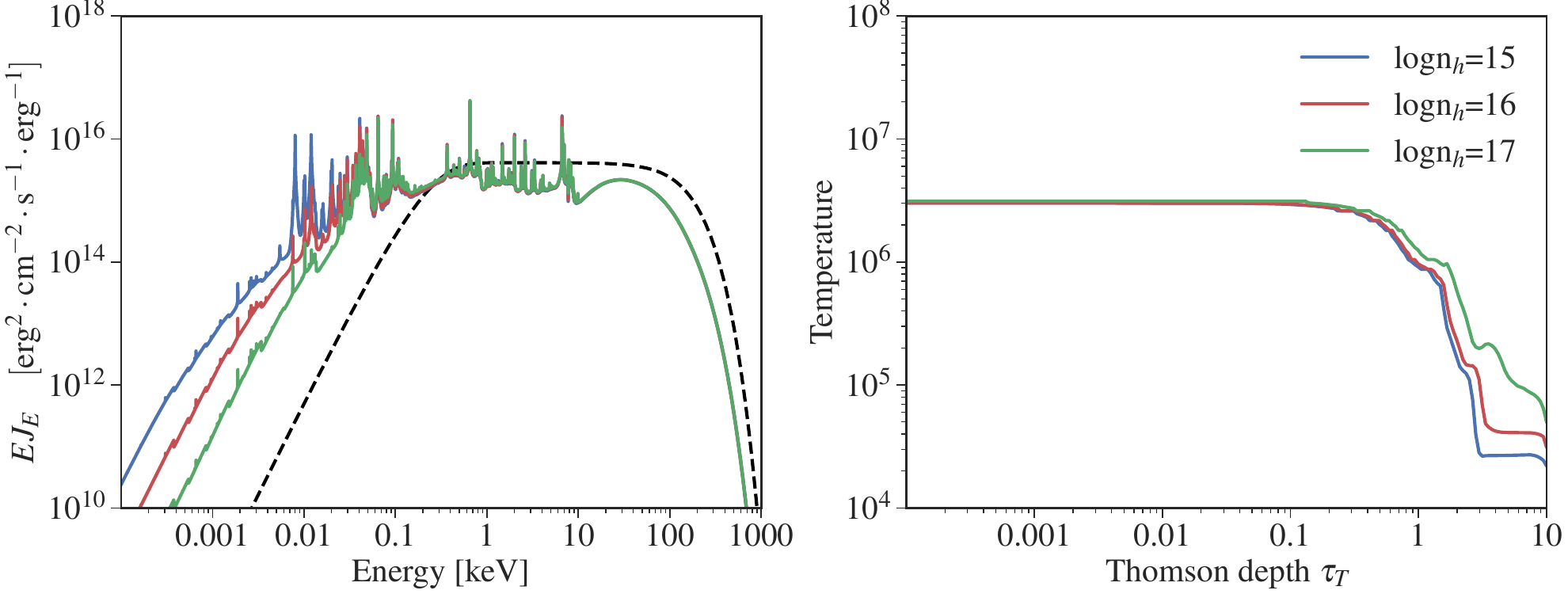}
    \caption{Angle-averaged reflection spectra for different hydrogen densities, normalized relative to the spectrum at $\log n_\mathrm{h} = 15$. Since the ionizing flux $F_x$ is proportional to the hydrogen density $n_\mathrm{h}$ (Eq.~\ref{eq:ionization par}), the spectra for $\log n_\mathrm{h} = 16$ and 17 are scaled by factors of 10 and 100, respectively. All other parameters are the same as the default values listed in Table~\ref{model parameters}.}
    \label{fig:nthcomp dif density}
\end{figure*}
\begin{figure*}[htbp]
    \centering
    \includegraphics[width=\linewidth]{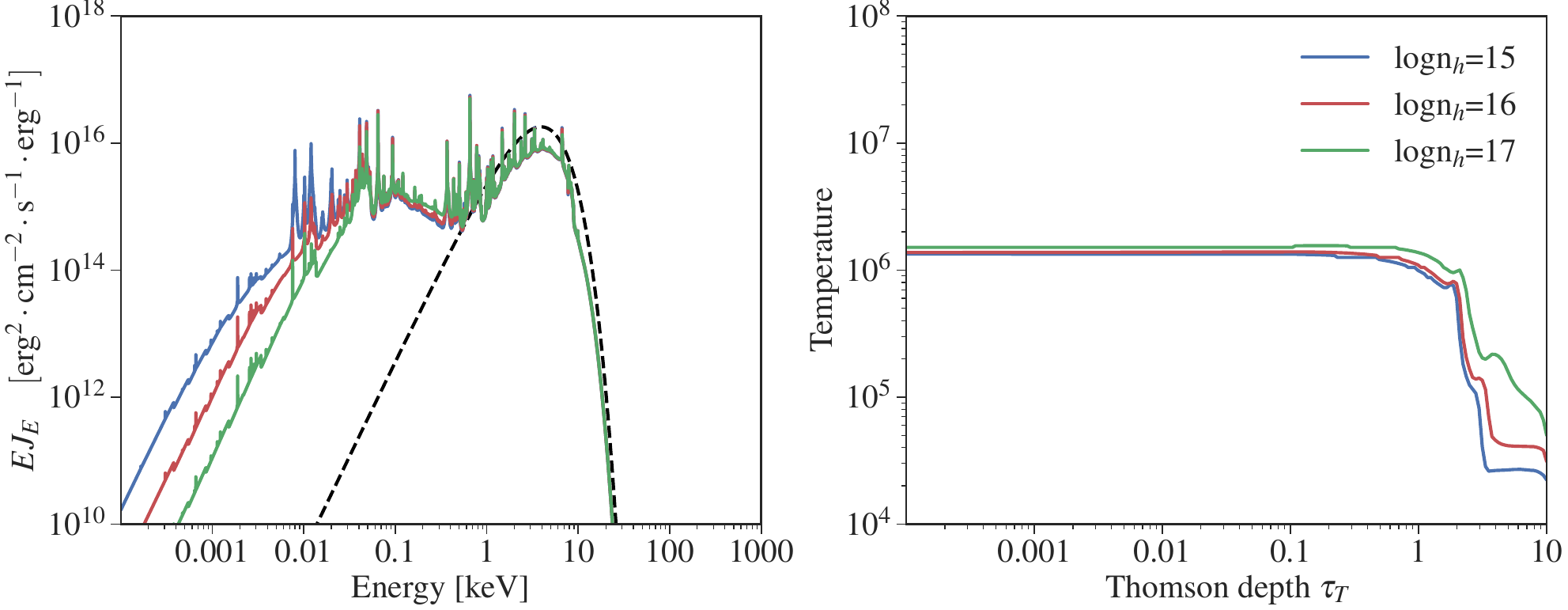}
    \caption{Consistent with the setup in Figure~\ref{fig:nthcomp dif density}, the incident radiation in this plot is modeled as a blackbody with $kT_\mathrm{bb} = 1$~keV.}
    \label{fig:blackbody dif density}
\end{figure*}
\begin{figure*}[htbp]
    \centering
    \includegraphics[width=\linewidth]{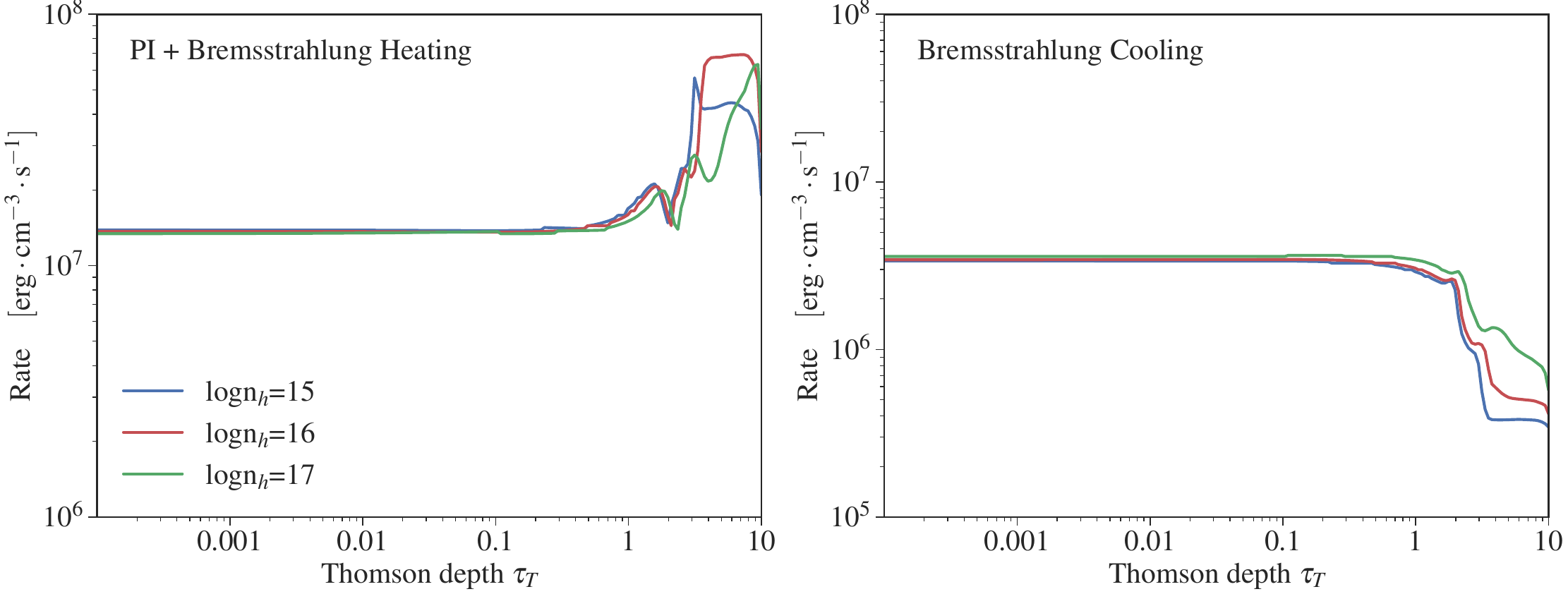}
    \caption{Photoionization and bremsstrahlung heating rates, along with the bremsstrahlung cooling rate, corresponding to the reflection spectra shown in Figure~\ref{fig:blackbody dif density}.}
    \label{fig:total heating-cooling}
\end{figure*}

\begin{figure*}[htbp]
    \centering
    \includegraphics[width=\linewidth]{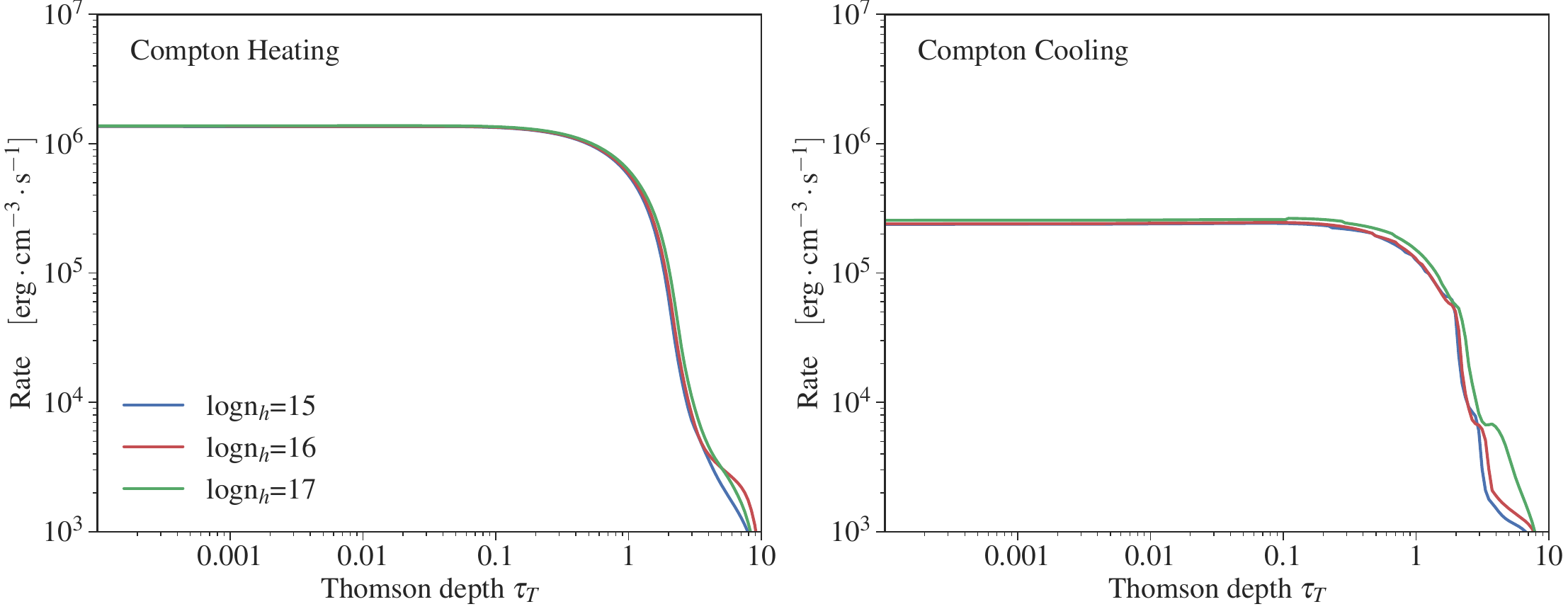}
    \caption{Compton heating and cooling rates, corresponding to the reflection spectra shown in Figure~\ref{fig:blackbody dif density}.}
    \label{fig:Compton heating-cooling}
\end{figure*}

\subsection{Effects of varying the hydrogen density}\label{chap:density}

Currently, \texttt{DAO} assumes reflection in a constant-density slab, with a default hydrogen density of 10$^{15}$ cm$^{-3}$. In AGN or XRB systems, the disk densities are sometimes greater than 10$^{15}$ cm$^{-3}$ \citep{2019MNRAS.489.3436J} and sometimes even higher than 10$^{20}$ cm$^{-3}$ \citep{2023ApJ...951..145L}. \cite{2016MNRAS.462..751G} and \cite{2024ApJ...974..280D} showed that the emergent reflection spectrum depends sensitively on the assumed disk density. However, the standard {\tt XSTAR} atomic table (ADTB) only supports hydrogen densities lower than 10$^{18}$ cm$^{-3}$; a specific atomic database is needed for higher hydrogen densities \citep{2021ApJ...908...94K}. \texttt{DAO} will be updated to consider higher densities once the atomic database is available. 

In Figure~\ref{fig:nthcomp dif density}, we compare reflection spectra for hydrogen densities of 10$^{15}$, 10$^{16}$, and 10$^{17}$ cm$^{-3}$ under \texttt{nthcomp} illumination. Other parameters are set at their default values. The results are consistent with the findings of \cite{2016MNRAS.462..751G}: for a given ionization parameter, the gas temperature increases with hydrogen density, and the continuum emission in the low energy band displays clear differences among these spectra. 

To further study the density effects in the soft X-ray band and temperature profile, we perform additional calculations with a blackbody incident spectrum of temperature 1~keV, which allows us to focus on the soft X-ray band. As shown in Figure~\ref{fig:blackbody dif density}, the Compton hump is weaker under this type of illumination than for a power-law incident spectrum, and the high-energy parts of the spectra become nearly identical across different densities. However, noticeable differences persist in the low-energy continuum emission. The temperature profiles show obvious differences at large optical depths. Our findings agree with those of \cite{2022ApJ...926...13G}, who introduced the \texttt{xillverNS} model that assumes a blackbody function as the illuminating spectrum.

The $n_{\rm h}^2$ dependence of bremsstrahlung heating and cooling rates plays an important role in shaping the temperature profile of the reflection region. The bremsstrahlung heating rate is related to the free--free absorption coefficient through
\begin{equation}
    H_{\rm ff} = \int \alpha_{\rm ff}(E) j(E) dE,
\end{equation}
where the free--free absorption coefficient $\alpha_{\rm ff}$ is
\begin{equation}
    \alpha_\nu^{\mathrm{ff}} = 3.7 \times 10^{8} Z^{2} n_{e} n_{z} \bar{g}_{\mathrm{ff}} T^{-1/2} \nu^{-3} \left( 1 - e^{-h\nu/kT} \right).
\end{equation}
Here, $n_e$ is the electron density, $n_z$ is the density of that ion, $\bar{g}_{\mathrm{ff}}(T,\nu)$ is the velocity-averaged Gaunt factor, $T$ is the electron temperature, $Z$ is the charge of the most abundant ion, and the bremsstrahlung emissivity $j$ is given by
\begin{equation}
    j_\nu = 6.8\times10^{-38} Z^2 n_e n_z T^{-1/2} e^{-h\nu/kT} \bar{g}_{\mathrm{ff}}.
\end{equation}
The bremsstrahlung cooling rate can be expressed as
\begin{equation}
    \Gamma_{\rm ff} = 1.42\times 10^{-27} T^{1/2} Z^2 n_e n_z \bar{g}_{ff}.
\end{equation}

It is evident from the above expressions that both the heating and cooling rates scale as $n_{\rm h}^2$, so even modest increases in density can dramatically alter the thermal balance of the gas. By contrast, the net Compton heating--cooling rate in the non-relativistic limit is given by Eq.\ref{eq:Compton heating}, which mainly depends on photons' energies. Thus, while Compton processes dominate in the hard X-ray band, bremsstrahlung heating and cooling become increasingly important in the high-density regime due to their quadratic dependence on $n_{\rm h}$.

Figures~\ref{fig:total heating-cooling}--\ref{fig:Compton heating-cooling} show the heating and cooling rates associated with the spectra in Figure~\ref{fig:blackbody dif density}. These rates govern the gas temperature at each layer and ultimately shape the emergent spectrum. As seen in Figure~\ref{fig:total heating-cooling}, bremsstrahlung heating and cooling vary strongly in the deeper layers. By contrast, Compton heating remains nearly independent of hydrogen density in these regions, and Compton cooling exhibits only minor variations compared with bremsstrahlung cooling. A larger discrepancy is expected for $\log(n_\mathrm{h}) > 18$, a regime that will be investigated in future work.

\section{discussion}\label{sec:discuess}

In this paper, we presented a new reflection model, which we named \texttt{DAO}. This model is open-source and highly flexible for studying the reflection phenomenon in accretion-disk-corona systems. The model assumes a slab geometry, with the parameters of the slab and the incident radiation field listed in Table \ref{model parameters}. Following the mathematical framework of the \texttt{xillver} models, \texttt{DAO} uses the {\tt XSTAR} code to calculate the photoionization of each layer and the Feautrier method to solve the radiative transfer through the layers.

We compared our new X-ray reflection model \texttt{DAO} with \texttt{xillver} and \texttt{reflionx}. Both \texttt{DAO} and \texttt{xillver} include more complete atomic databases, and a better treatment of angular dependence, than \texttt{reflionx}. \texttt{DAO} includes several features that were recently included in the proprietary \texttt{xillver} code, but have not yet been released as publicly available \texttt{xillver} table models. First, \texttt{DAO} includes the most complete and up-to-date atomic database available. This was explored for the \texttt{xillver} model by \cite{2024ApJ...974..280D} but has not yet been included in a publicly available table model. \texttt{DAO} implements an accurate treatment of Compton scattering, which was explored for the \texttt{xillver} model by \cite{2020ApJ...897...67G} but again has not yet been publicly released in a table model. We also include some effects that, to our knowledge, are not considered in \texttt{xillver}, such as the inclusion of nickel and employing a slightly more advanced treatment of the incidence angle. However, we believe that the main advantage of \texttt{DAO} is that it is open-source and allows users to freely define their own incident spectrum. This flexibility is crucial for several applications—for example, modelling returning radiation \citep{2024ApJ...976..229M, 2024ApJ...965...66M}, or describing reflection in Z sources where the illuminating spectrum is neither a pure power law nor a blackbody, but instead a saturated Comptonization continuum \citep{Ludlam2024Ap&SS.369...16L}.


In the future, we plan to extend our model to higher hydrogen densities to better represent environments with dense gas in XRBs and AGNs \citep[e.g.,][]{2019MNRAS.489.3436J, 2023ApJ...951..145L}, and to incorporate disk-blackbody radiation from the bottom boundary, which is particularly important for X-ray binaries \citep[e.g.,][]{2007MNRAS.381.1697R}. We will also implement a non-uniform vertical density structure and account for hydrostatic equilibrium \citep{2007MNRAS.377L..59D}. Another major development will be the inclusion of polarization in the radiative transfer calculations, including Compton scattering for polarized X-ray reflection spectra. We further intend to examine the impact of different atomic databases and photoionization codes, for example by integrating CLOUDY \citep{Gunasekera2025arXiv250801102G} into our framework.




{\bf Acknowledgments --}
We sincerely thank Ekaterina Sokolova-Lapa and Yuanze Ding for their insightful comments and generous assistance throughout the preparation of this paper.
 This work was supported by the National Natural Science Foundation of China (NSFC), Grant No.~W2531002. HL acknowledges support from The Gruber Foundation fellowship. AI acknowledges support by the European Union (ERC, X-MAPS, 101169908). Views and opinions expressed are however those of the author(s) only and do not necessarily reflect those of the European Union or the European Research Council. Neither the European Union nor the granting authority can be held responsible for them.

\bibliographystyle{aa}
\bibliography{sample701}

\appendix

\section{Algorithm}\label{apd:algorithm}

\noindent The second-order radiative transfer equation is given by
\begin{equation}\label{eq:secorder in APD}
    \mu^{2} \frac{\partial^2 u_{\mu\nu}}{\partial\tau^{2}_{\nu}} = u_{\mu\nu} + S_{\nu}
\end{equation}
The derivatives with respect to optical depth can be approximated using finite differences as
\begin{equation}
    \left(\frac{dX}{d\tau}\right)_{d-\frac{1}{2}} = \frac{X_d-X_{d-1}}{\tau_d-\tau_{d-1}},\quad    \left(\frac{dX}{d\tau}\right)_{d+\frac{1}{2}} = \frac{X_{d+1}-X_{d}}{\tau_{d+1}-\tau_{d}}
\end{equation}
and the second derivative at depth index $d$ is approximated by
\begin{equation}
    \left(\frac{d^2X}{d\tau^2}\right)_d \approx\frac{\left(dX/d\tau\right)_{d+\frac{1}{2}}-\left(dX/d\tau\right)_{d-\frac{1}{2}}}{\frac{1}{2}\left(\Delta\tau_{d+\frac{1}{2}}+\Delta\tau_{d-\frac{1}{2}}\right)}
\end{equation}
In this discretization, the derivative operator is applied to the spatial grid, while the energy and angle dependencies are treated implicitly. Defining the optical depth intervals as $\Delta\tau_{d+1/2} = \tau_{d+1}-\tau_d$ and $\Delta\tau_{d-1/2} = \tau_{d}-\tau_{d-1}$, the centered interval is given by
\begin{equation}
    \Delta\tau_d = \frac{1}{2} \left(\Delta\tau_{d-\frac{1}{2}} + \Delta\tau_{d+\frac{1}{2}}\right)
\end{equation}
Let $m$ and $n$ denote the angle and energy indices, respectively. The discrete form of Eq.~\ref{eq:secorder in APD} within the domain defined by $\tau_{\mathrm{min}}<\tau<\tau_{\mathrm{max}}$, $\mu_{\mathrm{min}}\leq\mu\leq\mu_{\mathrm{max}}$, and $\nu_{\mathrm{min}}\leq\nu\leq\nu_{\mathrm{max}}$ is expressed as:
\begin{equation}\label{eq:dif from rte}
    \frac{\mu_m^2u_{d-1,m,n}}{\Delta\tau_{d-\frac{1}{2},n}\Delta\tau_{d,n}} - \frac{\mu_m^2u_{d,m,n}}{\Delta\tau_{d,n}}\left(\frac{1}{\Delta\tau_{d-\frac{1}{2},n}}+\frac{1}{\Delta\tau_{d+\frac{1}{2},n}}\right) + \frac{\mu_m^2u_{d+1,m,n}}{\Delta\tau_{d+\frac{1}{2},n}\Delta\tau_{d,n}} = u_{d,n}-S_{d,n}
\end{equation}
At the upper boundary, the intensity $u$ at the second depth point ($d = 2$) can be expanded using a Taylor series around the surface ($d = 1$):
\begin{equation}
u_{2} = u_{1} + \Delta\tau_{\frac{3}{2}} \left(\frac{du}{d\tau}\right)_{d=1} + \frac{1}{2}\Delta\tau_{\frac{3}{2}}^{2} \left(\frac{d^{2}u}{d\tau^{2}}\right)_{d=1}
\end{equation}
Substituting this expansion into Eq.~\ref{eq:secorder in APD} yields
\begin{equation}\label{eq:bounady top apd}
\mu \frac{u_{2} - u_{1}}{\Delta\tau_{1}} = u_{1} + \Delta\tau_{1} \frac{u_{1} - S_{1}}{2\mu} + I_{\mathrm{inc}}.
\end{equation}
A similar procedure is applied to the lower boundary. Combining these boundary conditions with Eqs.~\ref{eq:dif from rte} and \ref{eq:bounady top apd} results in a block tridiagonal system of the form
\begin{equation}\label{eq:mati}
    -A_{d}u_{d-1} + B_du_d-C_du_{d+1} = R_d,\quad 2\le d\le ND-1
\end{equation}
In Eq.~\ref{eq:mati}, the subscripts denoting angle and energy are omitted for clarity. At the two boundaries, $d = 1$ and $d = ND$, the coronal illumination $I_{\mathrm{inc}}$ and the disk self-emission $I_{\mathrm{bb}}$ must be included in the source terms. Generally, $I_{\mathrm{inc}}$ is anisotropic and depends on the physical properties of the corona, while $I_{\mathrm{bb}}$ is an isotropic, multi-temperature blackbody spectrum. Eq.~\ref{eq:mati} can be reformulated in a compact vector notation as:
\begin{equation}\label{eq:comp mati}
\mathbf{T} \cdot \mathbf{u} = \mathbf{R},
\end{equation}
where $\mathbf{T}$ represents the coefficient matrix, $\mathbf{u}$ is the vector of mean intensities, and $\mathbf{R}$ is the corresponding source vector. Equation~\ref{eq:comp mati}, together with the boundary condition in Eq.~\ref{eq:bounady top apd}, is solvable via a recursive forward-elimination and back-substitution procedure \citep{1964CR....258.3189F,1967AnAp...30..125F,1968AnAp...31..257F}. This solution scheme is collectively known as the Feautrier method.

\begin{figure}
    \centering
    \includegraphics[width=0.8\linewidth]{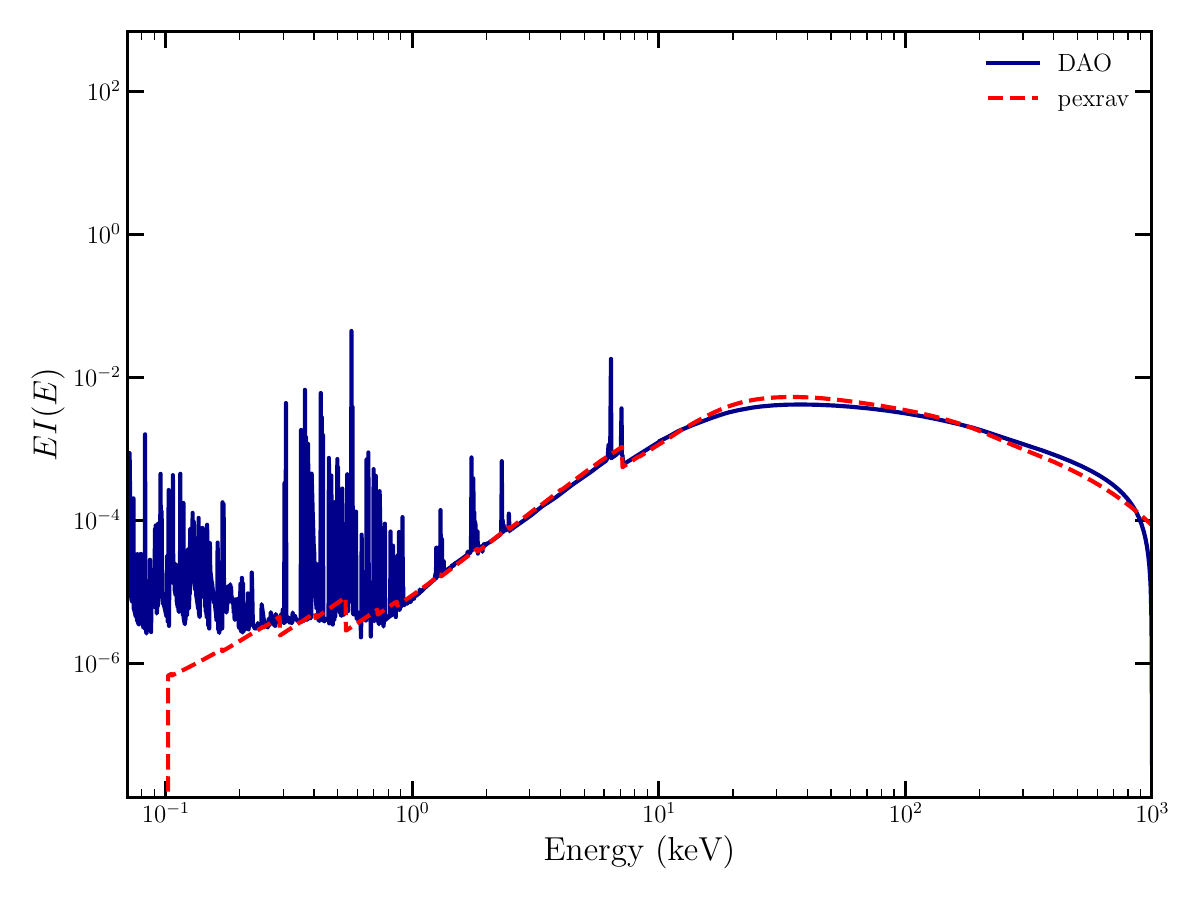}
    \caption{The illuminating continuum is a cut-off power law with a high-energy cut-off of $E_{\mathrm{cut}}=1000\ \mathrm{keV}$. For the $\texttt{DAO}$ model, the material is set to be neutral by fixing the ionization parameter at $\log\xi=0.0$. We adopt a standard iron abundance of $A_{\mathrm{Fe}} = 1.0$ and set the abundances of nickel, calcium, and argon to zero ($A_{\mathrm{Ni}} = 0.0$, $A_{\mathrm{Ca}} = 0.0$, $A_{\mathrm{Ar}} = 0.0$) to expedite the calculation. The other parameters in \texttt{DAO} are set at default values.}
    \label{fig:pexrav}
\end{figure}

We validate our radiative transfer code by comparing its results with the established reflection model $\texttt{pexrav}$ \citep{1995MNRAS.273..837M}, as illustrated in Figure \ref{fig:pexrav}. The spectra presented in Figure \ref{fig:pexrav} demonstrate excellent agreement in the continuum below $10\ \mathrm{keV}$. While the $\texttt{pexrav}$ spectrum exhibits a slightly higher peak in the Compton Hump range, the two models agree well at energies $E < 20\ \mathrm{keV}$ and $E > 50\ \mathrm{keV}$. We hypothesize that this difference will decrease upon the inclusion of the angle-dependent source function, as described in Section \ref{sec:METHOD}. It must be noted that the $\texttt{pexrav}$ model does not incorporate line emission; however, this exclusion is negligible when focusing on the high-energy scattered continuum.

\section{Compton scattering redistribution function}\label{apd:redistribution function}
\begin{figure*}
    \centering
    \includegraphics[width=\linewidth]{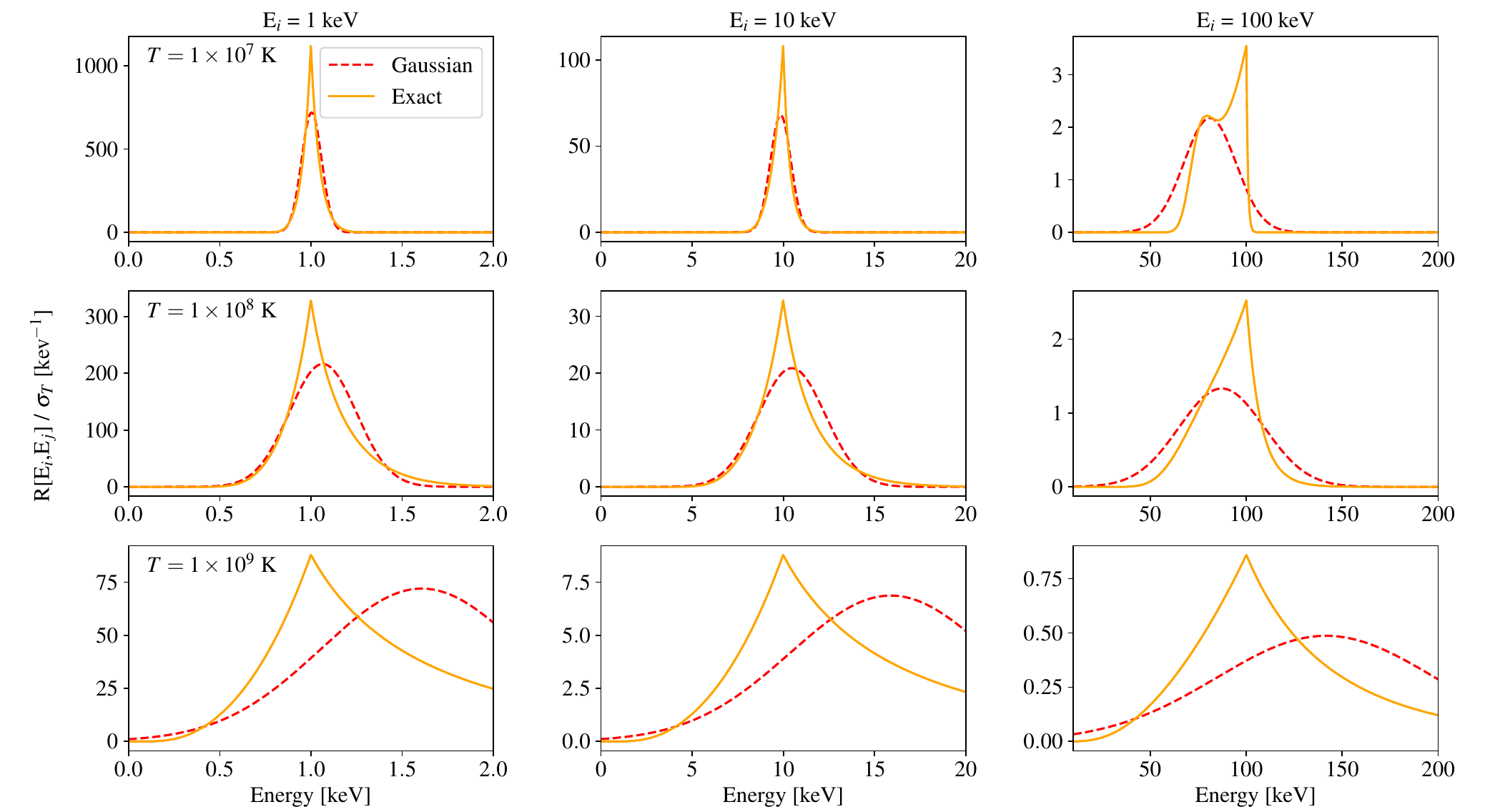}
    \caption{Gaussian (red dash line) and Exact redistribution (orange solid line) function with three initial energy: E$_i$ = 1, 10, 100 keV and three different gas temperature: 10$^7$, 10$^8$, 10$^9$ K.}
    \label{fig:kernel}
\end{figure*}

\subsection{Gaussian approximation redistribution function}

The Gaussian redistribution function was first applied to the scattering problem in Compton-thick atmospheres by \citet{1978ApJ...219..292R,1993MNRAS.261...74R} to describe the down-scattering of photons with $E < 200$~keV, and was later extended by \citet{2000ApJ...537..833N} to treat Compton scattering over the full energy range.

This redistribution function describes the probability that a photon with initial energy $E_i$ is scattered to energy $E_f$:
\begin{equation}
P(E_i,E_f) = \frac{1}{\sqrt{2\pi}\Sigma}
\exp\left[-\frac{(E_f-E_c)^2}{2\Sigma^2}\right],
\end{equation}
with the central energy $E_c$ and standard deviation $\Sigma$ defined as
\begin{equation}
E_c = E_i\left(1 + \frac{4kT}{m_ec^2} - \frac{E_i}{m_ec^2}\right),
\end{equation}
and
\begin{equation}
\Sigma = E_i\left[\frac{2kT}{m_ec^2} + \frac{2}{5}\left(\frac{E_i}{m_ec^2}\right)^2\right]^{1/2}.
\end{equation}

\subsection{Exact Redistribution Function}
Following the study of \citet{2020ApJ...897...67G}, 
we use the method described in \citet{1993AstL...19..262N,2017MNRAS.469.2032M} to calculate the exact redistribution function
\begin{equation}\label{eq:exact}
R_E(x,x_1,\mu,\gamma)= \frac{2}{Q} + \frac{u}{v} \left(1-\frac{2}{q}\right) + u \frac{(u^2-Q^2)(u^2+5v)}{2q^2v^3}+u\frac{Q^2}{q^2 v^2},
\end{equation}
where $x$ and $x_1$ represent the dimensionless photon energies before and after scattering, respectively, $\mu$ is the cosine of the scattering angle, and $\gamma$ denotes the electron Lorentz factor. Here, $q = xx_1 (1-\mu)$ and $Q^2=(x-x_1)^2+2q$, while the auxiliary variables are defined as
\begin{equation}
a_-^2 = (\gamma-x)^2 + \frac{1+\mu}{1-\mu}, \qquad
a_+^2 = (\gamma+x_1)^2 + \frac{1+\mu}{1-\mu},
\end{equation}
and
\begin{equation}
v = a_-a_+, \qquad
u = a_+ - a_-.
\end{equation}
The redistribution function averaged over a relativistic Maxwellian distribution is given by
\begin{equation}
R(x,x_1,\mu) = \frac{3}{32\mu\Theta K_2(1/\Theta)}\int _{\left(x-x_1+Q\sqrt{1+2/q}\right)/2}^\infty R_E(x,x_1,\mu,\gamma)\exp(-\gamma/\Theta)\,d\gamma
\end{equation}
where $K_2$ is the modified Bessel function of the second kind and $\Theta=kT/m_ec^2$. In this work, the angle-averaged redistribution is employed; thus, the resulting function is
\begin{equation}\label{eq:kernel_intmu}
    R(x,x_1) = \int_{\mu_{\mathrm{min}}}^{\mu_\mathrm{max}}R(x,x_1,\mu)\,d\mu
\end{equation}

Figure~\ref{fig:kernel} illustrates the significant deviation between the exact redistribution function and the Gaussian approximation when the gas temperature is high and the photon energies are large. Similar results were reported by \citet{2020ApJ...897...67G}; the comparison is reproduced here to facilitate the reader's understanding of the differences. The redistribution function in Eq.~\ref{eq:kernel_intmu} satisfies the normalization condition with respect to the Compton scattering cross section:
\begin{equation}
    \frac{\sigma_{\mathrm{cs}}(E)}{\sigma_T} = \int_0^\infty R(E_i,E_f)\,dE_f.
\end{equation}
Rather than the Klein--Nishina cross section, $\sigma_{\mathrm{KN}}$, the Compton scattering cross section averaged over a relativistic Maxwellian electron distribution, denoted by $\sigma_{\mathrm{cs}}$, is employed:
\begin{equation}\label{eq:rel_cs}
\begin{aligned}
\sigma_{\mathrm{cs}}(x) =
\frac{3\sigma_T}{16x^2\Theta K_2(1/\Theta)}
\int_1^\infty e^{-\gamma/\Theta}
\Biggl\{ &
\left(x\gamma+\frac{9}{2}+\frac{2\gamma}{x}\right)
\ln{\left[\frac{1+2x(\gamma+z)}{1+2x(\gamma-z)}\right]} - 2xz \\
& + z\left(x-\frac{2}{x}\right)\ln(1+4x\gamma+4x^2)
+\frac{4x^2z(\gamma+x)}{1+4x\gamma+4x^2} \\
& -2\int_{x(\gamma-z)}^{x(\gamma+z)}
\ln(1+2\varepsilon)\frac{d\varepsilon}{\varepsilon}
\Biggr\} d\gamma.
\end{aligned}
\end{equation}
A detailed comparison between $\sigma_{\mathrm{cs}}$ and $\sigma_{\mathrm{KN}}$ is provided in \citet{2020ApJ...897...67G}.

\section{Relativistic spectrum}\label{apd:rel}
The \texttt{reldao} and \texttt{reldaoA} models are developed within the framework of \texttt{relxill}~v2.5 \citep{2014ApJ...782...76G,2025ApJ...989..168H} and \texttt{relxillA} \citep{2025ApJ...989..168H}. Currently, as the full table models have not yet been generated, \texttt{reldao} serves as a prototype rather than a finalized relativistic reflection model. The code will be publicly released upon the completion of the \texttt{DAO} table models. Additionally, the relativistic spectrum is computed using the convolution model \texttt{relconv} \citep{2010MNRAS.409.1534D} within \texttt{XSPEC} \citep{1996ASPC..101...17A}, yielding the combined model \texttt{relconv*dao}.

The primary distinction between the convolution model \texttt{relconv*dao} and the direct implementations, \texttt{reldao} and \texttt{reldaoA}, lies in the treatment of angular dependence. The convolution model operates on the angle-averaged flux, whereas \texttt{reldao} accounts for the full angular distribution by integrating over all local emission angles at a given incidence angle. Furthermore, \texttt{relxillA} incorporates a more comprehensive treatment of angular effects. In the specific case of \texttt{reldao}, the observed flux is calculated as
\begin{equation}
    F_{obs} (E_{obs}) = \frac{1}{D^2} \int_{R_{in}}^{R_{out}} dr_e \int_0^1 g^* \frac{\pi r_e g^2}{\sqrt{g^*(1-g^*)}} \left[f^{(1)}(g^*,r_e,\theta_{obs})+f^{(2)}(g^*,r_e,\theta_{obs})\right] \times \epsilon(r_e) \langle\bar{I_e}(E_e)\rangle
\end{equation}
and for \texttt{reldaoA}, the model calculates the observed flux as 
\begin{equation}
\begin{aligned}
        F_{obs} (E_{obs}) &= \frac{1}{D^2} \sum_{i=0}^9 \int_{R_{in}}^{R_{out}}dr_e\int_0^1 dg^* \frac{\pi r_eg^2}{\sqrt{g^*(1-g^*)}}\left[f^{(1)}(g^*,r_e,\theta_{obs})+f^{(2)}(g^*,r_e,\theta_{obs})\right]  \\
        &\times\epsilon(r_e)\bar{I}(E_e,r_e,\bar{\theta_e})\Theta(\theta_e-\theta_i)\Theta(\theta_{i+1-\theta_e})
\end{aligned}
\end{equation}
All the physical parameters above have been explained in detail in section 2 of \cite{2025ApJ...989..168H}.
\begin{figure*}
    \centering
    \includegraphics[width=\linewidth]{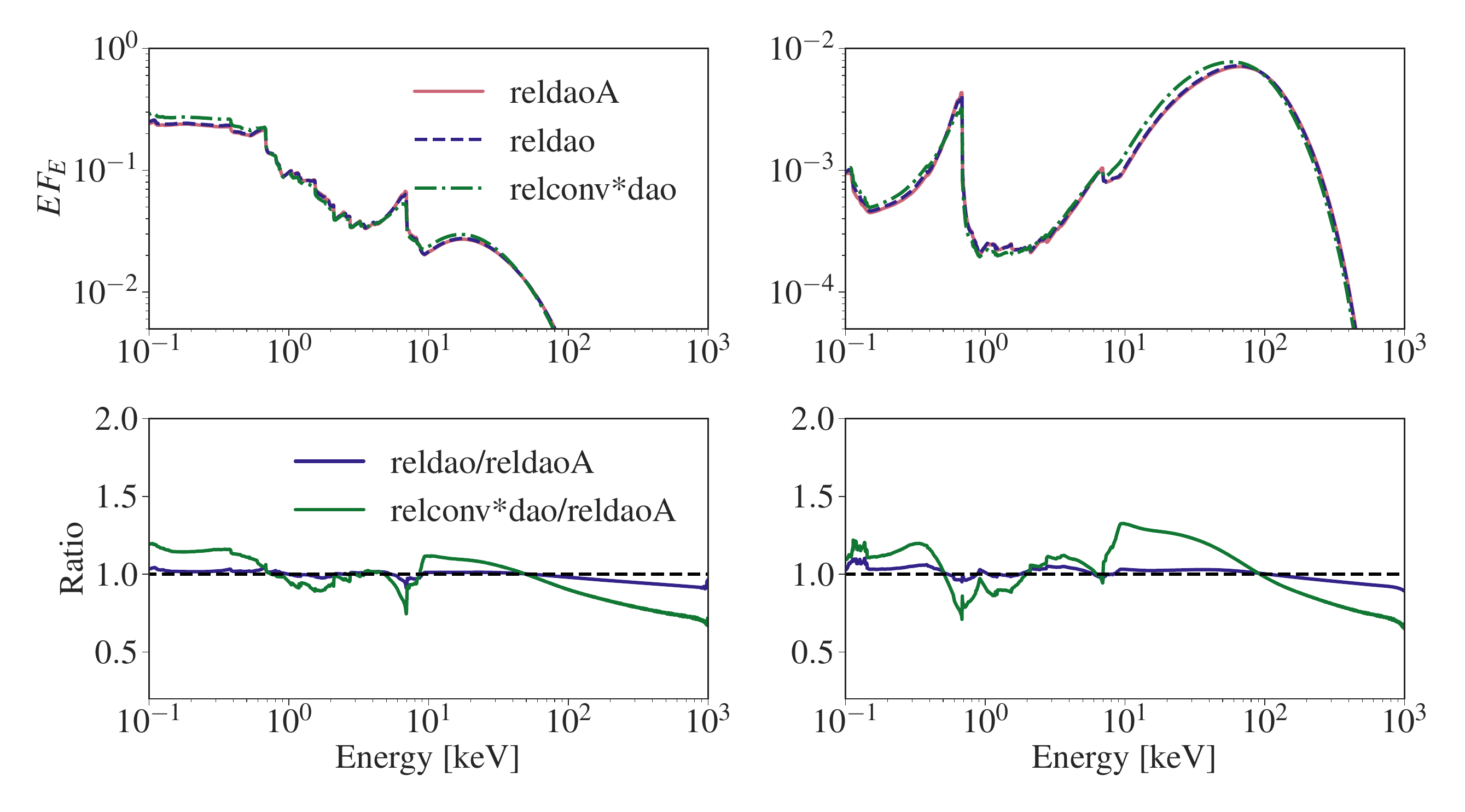}
    \caption{Relativistic reflection spectra calculated with \texttt{relconv*dao} (green dash-dot line), \texttt{reldao} (blue dashed line), and \texttt{reldaoA} (red solid line). The gas is illuminated by \texttt{nthcomp} with $\Gamma$ = 2.4 (left column) and $\Gamma$ = 1.4 (right column). Ionization parameter is set to 3.0 for both cases. The inclination angle is $\theta_{i}$ = 30 deg, and all other parameters in \texttt{relxill} are fixed at their default values.}
    \label{fig:reldao}
\end{figure*}
The relativistic reflection spectra are presented in Figure~\ref{fig:reldao}. Each spectrum is normalized by its total flux, according to $F(E) = F(E) / \int F(E) \, dE$. The \texttt{reldao} and \texttt{reldaoA} models yield consistent results, whereas the convolution model, \texttt{relconv*dao}, exhibits marked deviations in both the Compton hump region and the soft X-ray band.




\end{document}